# The Complexity of Computing the Size of an Interval[1]

Lane A. Hemaspaandra
Department of Computer Science
University of Rochester
Rochester, NY 14627
USA

Christopher M. Homan[2]
Department of Computer Science
Rochester Institute of Technology
Rochester, NY 14623
USA

Sven Kosub[3]
Institut für Informatik
Technische Universität München
D-85748 Garching b. München
Germany

Klaus W. Wagner[4]
Institut für Informatik
Julius-Maximilians-Univ. Würzburg
D-97074 Würzburg
Germany

February 13, 2005; revised March 16, 2005

[1]A preliminary version of some parts of this paper was presented at the 28th International Colloquium on Automata, Languages and Programming held in Crete, Greece, in July 2001 [HKW01]. Supported in part by grants NSF-CCR-9322513, NSF-INT-9815095/DAAD-315-PPP-gü-ab, and NSF-CCF-0426761. Work done in part while the second author was at the University of Rochester, and in part while the first two authors were visiting Julius-Maximilians-Universität Würzburg.

[2]Email: `cmh@cs.rit.edu`.
[3]Email: `kosub@in.tum.de`.
[4]Email: `wagner@informatik.uni-wuerzburg.de`.

**Abstract**

Given a p-order $A$ over a universe of strings (i.e., a transitive, reflexive, antisymmetric relation such that if $(x, y) \in A$ then $|x|$ is polynomially bounded by $|y|$), an interval size function of $A$ returns, for each string $x$ in the universe, the number of strings in the interval between strings $b(x)$ and $t(x)$ (with respect to $A$), where $b(x)$ and $t(x)$ are functions that are polynomial-time computable in the length of $x$.

By choosing sets of interval size functions based on feasibility requirements for their underlying p-orders, we obtain new characterizations of complexity classes. We prove that the set of all interval size functions whose underlying p-orders are polynomial-time decidable is exactly #P. We show that the interval size functions for orders with polynomial-time adjacency checks are closely related to the class FPSPACE(poly). Indeed, FPSPACE(poly) is exactly the class of all nonnegative functions that are an interval size function minus a polynomial-time computable function.

We study two important functions in relation to interval size functions. The function #DIV maps each natural number $n$ to the number of nontrivial divisors of $n$. We show that #DIV is an interval size function of a polynomial-time decidable partial p-order with polynomial-time adjacency checks. The function #MONSAT maps each monotone boolean formula $F$ to the number of satisfying assignments of $F$. We show that #MONSAT is an interval size function of a polynomial-time decidable total p-order with polynomial-time adjacency checks.

Finally, we explore the related notion of cluster computation.

# 1 Introduction

The class NP, which is widely believed to contain computationally intractable problems, captures the complexity of determining for a given problem instance whether at least one suitable affirmative solution exists within an exponentially large set of (polynomial-sized) potential solutions. It is certainly not simpler, and seemingly much harder, to count all affirmative solutions in such solution sets. The corresponding *counting functions* constitute Valiant's widely studied counting class #P [Val79]. In the theory of counting functions, which is devoted to the study of counting versions of decision problems, most classes considered try to capture the pure phenomenon of counting, and in doing so they obscure other factors, e.g., orders on solution sets.

Natural counting problems in #P, of course, sometimes exhibit strong relationships between solutions to the problems. As an example, consider the counting function #DIV, which counts for each natural number the number of its nontrivial divisors. Clearly, #DIV is in #P since division can be done in polynomial time. A suitable structure in the set of solutions is the partial order of divisibility, that is, the order defined by $n \leq_{|} m$ iff $n$ divides $m$. Obviously, $\#\mathrm{DIV}(m) = \|\{k \mid 1 <_{|} k <_{|} m\}\|$, i.e., $\#\mathrm{DIV}(m)$ counts the number of elements in the open interval $(1, m)$ in the partial order "$\leq_{|}$" on natural numbers.

Is #DIV an exceptional case among #P functions in that it has such an interval size characterization? Interestingly, "no" is the answer. It turns out that a function $f$ is in #P if and only if it is an interval size function of a P-decidable partial p-order. The latter means that there exist a *partial p-order* $A$ (i.e., $A$ is a partial order and in addition satisfies the requirement that for some polynomial $p$ and all $x$ and $y$, it holds that $x \leq_A y$ implies $|x| \leq p(|y|)$) that is P-*decidable* (i.e., $x \leq_A y$ is decidable in polynomial time) and polynomial-time computable functions $b$ and $t$ such that $f(x) = \|\{z \mid b(x) <_A z <_A t(x)\}\|$, where $a <_A b$ denotes $a \leq_A b \wedge a \neq b$.

However, knowing that a partial p-order is polynomial-time decidable does not give us as much information as sometimes is needed. For example, the polynomial-time decidability of a p-order seemingly does not ensure that it has *efficient adjacency checks*, i.e., that there is a polynomial-time algorithm checking whether two elements are adjacent in this partial p-order. Indeed, if every P-decidable partial p-order has efficient adjacency checks then P = NP (and vice versa). Hence adding efficient adjacency checks to the properties listed above seems to be a restriction. Denote by $\mathrm{IF_p}$ the class of interval size functions of P-decidable partial p-orders with efficient adjacency checks. Denote by $\mathrm{IF_t}$ the class of interval size functions of P-decidable total p-orders with efficient adjacency checks. We have $\mathrm{IF_t} \subseteq \mathrm{IF_p} \subseteq \#\mathrm{P}$. Are these containments proper? On one hand, we prove that $\mathrm{IF_t} - \mathrm{FP} = \mathrm{IF_p} - \mathrm{FP} = \#\mathrm{P} - \mathrm{FP}$, where $A - B = \{a - b \mid a \in A \wedge b \in B\}$. Thus these three classes do not seem to be very different; indeed, they are identical given the smoothing power of subtracting polynomial-time computable adjustments. On the other hand, $\mathrm{IF_p} = \#\mathrm{P}$ is equivalent to P = NP, and $\mathrm{IF_t} = \mathrm{IF_p}$ only if UP = PH. Thus it is unlikely that any two of $\mathrm{IF_t}$, $\mathrm{IF_p}$, and #P coincide. Further, we study relationships between the classes $\mathrm{IF_t}$, FP, and $\mathrm{UPSV}_t$.

We already mentioned that it is unlikely that every P-decidable partial p-order has efficient adjacency checks. What about the converse? This also is not likely; if every partial p-order with efficient adjacency checks is P-decidable then P = PSPACE (and vice versa). Hence, in the presence of efficient adjacency checks, removing the P-decidability requirement seems to be a relaxation. Denote by $\mathrm{IF_p^*}$ the class of interval size functions of partial p-orders with efficient adjacency checks. Denote by $\mathrm{IF_t^*}$ the class of interval size functions of total p-orders with efficient adjacency checks. We have $\mathrm{IF_p} \subseteq \mathrm{IF_p^*}$ and $\mathrm{IF_t} \subseteq \mathrm{IF_t^*} \subseteq \mathrm{IF_p^*} \subseteq \mathrm{FPSPACE(poly)}$. We prove that $\mathrm{IF_t^*}$ (and $\mathrm{IF_p^*}$) are remarkably powerful: $\mathrm{IF_t^*} - \mathrm{FP} = \mathrm{FPSPACE(poly)} - \mathrm{FP}$. Thus $\mathrm{IF_t^*}$ (and $\mathrm{IF_p^*}$) are in a certain sense close to FPSPACE(poly), the class of polynomially length-bounded, polynomial-space computable functions. Nonetheless, we show that if these classes coincide, then UP = PSPACE. We clarify further relationships among such classes and also with respect to other function classes, in order to understand the power of interval computing.

We study two important natural functions in relation to interval size functions. We prove that the counting function #DIV is in $\mathrm{IF_p}$. Also, we show that the function #MONSAT, which counts for each monotone boolean formula the number of satisfying assignments that it has, belongs to $\mathrm{IF_t}$.

Using order-theoretic notions to approach complexity issues has a rich tradition, and appears in the literature in a variety of settings (e.g., [GHJY91, GS91, VW95, HVW96, Kos99]). The approaches in the examples just cited differ in intent from our approach in that they are based on a specific ordering, namely the lexicographical ordering. In contrast, for our purposes it is essential to consider more general feasible orderings (see [MP79, Ko83]).

Among earlier studies, perhaps the notion lying nearest to our approach is that of a cluster machine, which is a nondeterministic Turing machine that satisfies the promise that, on each input, all accepting computation paths are always neighbors with respect to the lexicographical ordering, i.e., the accepting paths must form a "cluster" [Kos99]. Based on this machine type, Kosub [Kos99] defined the counting class c#P (in a manner analogous to the way that #P is based on standard, nondeterministic polynomial-time Turing machines). Kosub obtained many interesting results about c#P, e.g., c#P seems to differ dramatically from #P in its closure properties (as regards, e.g., integer division, see [OH93, Kos99]), and he showed that c#P is closely related to a relatively simple unambiguous-nondeterminism-based function class, "$\mathrm{UPSV_t}$."

Most of the known results about c#P are proven by techniques that are exceedingly dependent on the fact that c#P is defined using adjacency clusters *with respect to lexicographic order*. In particular, the fact that in lexicographic order the function $f(a,b) = \|\{c \mid a \leq_{\mathrm{lex}} c \leq_{\mathrm{lex}} b\}\|$ is easy to compute underpins the results.

In the present paper we define the class CL#P, which studies the complexity of cluster computing in a context of relatively general (though length-respecting and having efficient adjacency checks) orders, rather than merely in the extremely special case of lexicographic order. We study CL#P and show, for example, that it does not equal c#P unless UP = PP (and thus the polynomial hierarchy collapses). On the other hand, we also prove that c#P and CL#P coincide on polynomially bounded functions, and that CL#P shows some behaviors quite reminiscent of c#P, e.g., though #P is closed under increment, we show that CL#P is closed under increment only if unexpected complexity collapses occur. More generally, we explore the relationship between CL#P and such classes as $\mathrm{IF_t}$, $\mathrm{IF_p}$, and #P. Though CL#P is in general flavor like an interval function (over a total order satisfying appropriate conditions but freed from the polynomial-time computability constraints of the functions defining the top and bottom of the interval), our results usually show that CL#P differs from the these classes unless unexpected complexity class collapses occur.

## 2 Preliminaries

Fix our finite alphabet to be $\Sigma = \{0,1\}$, and let $\Sigma^*$ denote the set of all finite strings over $\Sigma$. Let $\varepsilon$ denote the empty string. The length of a string $x \in \Sigma^*$ is denoted by $|x|$. The set of all strings of length $n$ is denoted by $\Sigma^n$. The complement of a set $L \subseteq \Sigma^*$ is denoted by $\overline{L}$, i.e., $\overline{L} = \Sigma^* \setminus L$. For any class $\mathcal{K}$ of subsets of $\Sigma^*$, let co$\mathcal{K}$ be the class $\{L \subseteq \Sigma^* \mid \overline{L} \in \mathcal{K}\}$. The cardinality of a finite set $S$ is denoted by $\|S\|$. The characteristic function of a set $L \subseteq \Sigma^*$ is denoted by $\chi_L$, i.e., for all $x \in \Sigma^*$, $\chi_L(x) = 1 \Leftrightarrow x \in L$ and $\chi_L(x) = 0 \Leftrightarrow x \notin L$. Let $\mathbb{N}$ denote the set $\{0,1,2,\ldots\}$. Let $\mathbb{N}^+$ denote the set $\{1,2,3,\ldots\}$.

For the basic notions of complexity theory such as P, NP, PSPACE, and so on see, e.g., the handbook [HO02].

The computation model we use is the standard nondeterministic Turing machine.

We review the definitions of some complexity classes of functions, already existing in the literature, that we will use in this paper.

- FP is the class of all (deterministic) polynomial-time computable, total functions from $\Sigma^*$ to $\mathbb{N}$. We will at times use FP to mean the class of all polynomial-time computable, total functions from $\Sigma^*$ to $\Sigma^*$. Via the natural, efficient bijection between $\mathbb{N}$ and $\Sigma^*$, these two notions are essentially the same.

- [Lad89] FPSPACE(poly) is the class of all polynomial-space computable, total functions from $\Sigma^*$ to $\mathbb{N}$ having polynomially length-bounded outputs. We will at times use FPSPACE(poly) to mean the class of all polynomial-space computable, total functions from $\Sigma^*$ to $\Sigma^*$ having polynomially length-bounded outputs. Via the natural, efficient bijection between $\mathbb{N}$ and $\Sigma^*$, these two notions are essentially the same.

- [Val79] #P is the class of all total functions $f$ for which there exists a nondeterministic polynomial-time Turing machine $M$ such that, for each $x$, $f(x)$ is the number of accepting computations of $M(x)$. Equivalently, #P is the class of all total functions $f$ for which there exist a set $B \in \mathrm{P}$ and a polynomial $p$ such that, for all $x \in \Sigma^*$, $f(x) = \|\{z \mid |z| = p(|x|) \wedge (x,z) \in B\}\|$.

- [GS88, Kos99] $\mathrm{UPSV_t}$ is the class of all total functions $f$ for which there exists a nondeterministic polynomial-time Turing machine $M$ that, on each input $x \in \Sigma^*$, has exactly one accepting path, and the output of this unique accepting path is $f(x)$.

For function classes $\mathcal{F}$ and $\mathcal{G}$ where each $f \in \mathcal{F} \cup \mathcal{G}$ maps from $\Sigma^*$ to $\mathbb{N}$, let $\mathcal{F} - \mathcal{G}$ denote the class of all functions $\{f - g \mid f \in \mathcal{F} \text{ and } g \in \mathcal{G}\}$. Note that the codomain of $\mathcal{F} - \mathcal{G}$ functions is $\{\ldots, -3, -2, -1, 0, 1, 2, 3, \ldots\}$. For each class $\mathcal{K}$ of sets, let $\mathrm{FP}^{\mathcal{K}}$ (respectively, $\mathrm{P}^{\mathcal{K}}$) be the class of functions (respectively, sets) that can be computed in polynomial time with an oracle from $\mathcal{K}$.

Next, we review the definitions of some complexity classes (of sets), already existing in the literature, that we will use in this paper.

- [Val76] UP is the class of all sets $L$ such that $\chi_L \in$ #P.

- [Coo71, Lev73] NP is the class of all sets $L$ for which there exists a function $f \in$ #P such that, for all $x \in \Sigma^*$, $x \in L \Leftrightarrow f(x) > 0$.

- [Sim75, Gil77] PP is the class of all sets $L$ for which there exist functions $f \in$ #P and $g \in$ FP such that, for all $x \in \Sigma^*$, $x \in L \Leftrightarrow f(x) \geq g(x)$.

- [OH93, FFK94] SPP is the class of all sets $L$ such that $\chi_L \in$ #P − FP.

- [CH90] Few is the class of all sets $L$ for which there exist a function $f \in$ #P, a set $B \in \mathrm{P}$, and a polynomial $p$ such that, for all $x \in \Sigma^*$, $f(x) \leq p(|x|)$ and $x \in L \Leftrightarrow (x, 1^{f(x)}) \in B$. In this definition, changing from "$f(x) \leq p(|x|)$" to "$0 < f(x) \leq p(|x|)$" can easily be seen to also yield Few.

- [MS72, Sto77] $\mathrm{PH} = \mathrm{P} \cup \mathrm{NP} \cup \mathrm{NP}^{\mathrm{NP}} \cup \mathrm{NP}^{\mathrm{NP}^{\mathrm{NP}}} \cup \ldots$ .

The following results are well-known or easy to see.

**Proposition 2.1** *1.* $\mathrm{FP} \subseteq \mathrm{UPSV_t} = \mathrm{FP}^{\mathrm{UP}\cap\mathrm{coUP}} \subseteq \#\mathrm{P} \subseteq \mathrm{FPSPACE(poly)}$.

*2.* $\mathrm{P} \subseteq \mathrm{UP} \subseteq \mathrm{Few} \cap \mathrm{NP} \subseteq \mathrm{Few} \cup \mathrm{NP} \subseteq \mathrm{P}^{\mathrm{NP}} \subseteq \mathrm{PH} \subseteq \mathrm{PSPACE}$.

*3.* $\mathrm{NP} \cup \mathrm{SPP} \subseteq \mathrm{PP}$.

*4.* [KSTT92] $\mathrm{Few} \subseteq \mathrm{SPP}$.

In this paper, we will sometimes for conciseness refer to the $j$th part of Theorem $i$ as Theorem $i.j$, e.g., we may refer to the third part of the above proposition as Proposition 2.1.3.

We will use the complexity-theoretic function-to-set operator $\exists$ of Hempel and Wechsung [HW00], which maps function classes to set classes. For a function class $\mathcal{F}$, $\exists \cdot \mathcal{F}$ is the class of all sets $L$ for which there exists a function $f \in \mathcal{F}$ such that, for all $x \in \Sigma^*$, $x \in L \Leftrightarrow f(x) > 0$.

The following statements are easy to see.

**Proposition 2.2** *1.* $\exists \cdot \mathrm{FP} = \exists \cdot (\mathrm{FP} - \mathrm{FP}) = \mathrm{P}$.

*2.* $\exists \cdot \mathrm{UPSV_t} = \exists \cdot (\mathrm{UPSV_t} - \mathrm{FP}) = \exists \cdot (\mathrm{UPSV_t} - \mathrm{UPSV_t}) = \mathrm{UP} \cap \mathrm{coUP}$.

*3.* $\exists \cdot \#\mathrm{P} = \mathrm{NP}$.

*4.* $\exists \cdot (\#\mathrm{P} - \mathrm{FP}) = \mathrm{PP}$.

*5.* $\exists \cdot \mathrm{FPSPACE(poly)} = \mathrm{PSPACE}$.

# 3 Orders with Feasibility Constraints

In this section, we define the notions of ordering that we use for the remainder of this paper (see also [Ko83]).

A binary relation $A \subseteq \Sigma^* \times \Sigma^*$ is a *partial order* if it is reflexive, antisymmetric (i.e., $(\forall x, y \in \Sigma^*)[x \neq y \implies ((x,y) \notin A \vee (y,x) \notin A)]$), and transitive. A partial order $A$ is a *total order* if, for all $x, y \in \Sigma^*$, $(x,y) \in A$ or $(y,x) \in A$. A partial order $A$ is a *partial p-order* if there exists a polynomial $q$ such that for all $(x,y) \in A$ it holds that $|x| \leq q(|y|)$.

For any partial p-order $A$, we employ the following standard notational conventions. We write $x \leq_A y$ if $(x,y) \in A$. We write $x <_A y$ if $x \leq_A y$ and $x \neq y$. We write $x \prec_A y$ if $x <_A y$ and there is no $z$ such that $x <_A z <_A y$. If $x \prec_A y$, we say that $x$ *precedes* $y$ or, equivalently, $y$ *succeeds* $x$. We let $A_\prec =_{\mathrm{def}} \{(x,y) \mid x \prec_A y\}$. The lexicographical order is denoted by $\leq_{\mathrm{lex}}$, and lexicographical adjacency is denoted by $\prec_{\mathrm{lex}}$.

Note that, for every partial p-order $A$ and every string $y$, there exist at most exponentially (in the length of $y$) many strings that are less than $y$ with respect to $A$. Thus, the output of an interval size function on a partial p-order is always at most exponential in the input length. Note that such exponential value bounds are typically the case with function classes, such as FP and #P, that are based on Turing machines having polynomial-time running bounds.

Feasibility constraints on orders are essential to our study. A partial p-order $A$ is P*-decidable* if $A \in \mathrm{P}$. A partial p-order $A$ is said to have *efficient adjacency checks* if $A_\prec \in \mathrm{P}$.

There are complexity-theoretic connections between these two feasibility requirements.

**Proposition 3.1** *Let $A$ be a partial p-order.*

1. *If* $A \in \mathrm{P}$ *then* $A_{\prec} \in \mathrm{coNP}$.

2. *If* $A_{\prec} \in \mathrm{P}$ *then* $A \in \mathrm{PSPACE}$.

*Proof.* The proof of (1) is immediate.

For (2), let $A$ be a partial p-order that has efficient adjacency checks. Let $M$ be an NPSPACE machine that accepts $A$ by, on input $(x, y)$, accepting immediately if $x = y$ and otherwise guessing a sequence $z_1, \ldots, z_k$ such that $x \prec_A z_1 \prec_A z_2 \prec_A \cdots \prec_A z_k \prec_A y$. Since $A$ is a partial p-order, for each $i \in \{1, \ldots, k\}$, $|z_i|$ is polynomially bounded with respect to $|y|$, so we only need guess such $z_i$'s whose lengths are polynomially bounded in $|y|$. So $A \in \mathrm{NPSPACE}$. However, as is well known, NPSPACE = PSPACE. ❑

**Corollary 3.2** 1. *If* P = NP*, then all* P*-decidable partial p-orders have efficient adjacency checks.*

2. *If* P = PSPACE*, then all partial p-orders with efficient adjacency checks are* P*-decidable.*

In what follows we will see that the converse of each of the claims of Corollary 3.2 also holds.

# 4 Orders without Efficient Adjacency Checks

We say that a function $f : \Sigma^* \to \mathbb{N}$ is an *interval size function* if there exist *boundary functions* $b$ and $t$ mapping from $\Sigma^*$ to $\Sigma^*$ and a partial order $A \subseteq \Sigma^* \times \Sigma^*$ such that, for all $x \in \Sigma^*$, $f(x) = ||\{z \mid b(x) <_A z <_A t(x)\}||$. In this section, we characterize #P in terms of interval size functions with polynomial-time decidable p-orders and polynomial-time computable boundary functions. We also note that if we omit all feasibility restrictions on p-orders, then all polynomially length-bounded functions can be characterized in a manner analogous to the way that interval size functions of resource-bounded orders characterize #P.

**Theorem 4.1** 1. *For any function $f$, the following statements are equivalent.*

   (a) $f \in \#\mathrm{P}$.

   (b) *There exist a partial p-order* $A \in \mathrm{P}$ *and functions* $b, t \in \mathrm{FP}$ *such that, for all* $x \in \Sigma^*$, $f(x) = \|\{z \mid b(x) <_A z <_A t(x)\}\|$.

   (c) *There exist a total p-order* $A \in \mathrm{P}$ *and functions* $b, t \in \mathrm{FP}$ *such that, for all* $x \in \Sigma^*$, $b(x) \leq_A t(x)$ *and* $f(x) = \|\{z \mid b(x) <_A z <_A t(x)\}\|$.

2. *For any function $f$ the following statements are equivalent.*

   (a) *$f$ is polynomially length-bounded.*

   (b) *There exist a partial p-order $A$ and functions* $b, t \in \mathrm{FP}$ *such that, for all* $x \in \Sigma^*$, $f(x) = \|\{z \mid b(x) <_A z <_A t(x)\}\|$.

   (c) *There exist a total p-order $A$ and functions* $b, t \in \mathrm{FP}$ *such that, for all* $x \in \Sigma^*$, $b(x) \leq_A t(x)$ *and* $f(x) = \|\{z \mid b(x) <_A z <_A t(x)\}\|$.

*Proof.* The implications (1c) ⇒ (1b), (1b)⇒ (1a), (2c) ⇒ (2b), and (2b) ⇒ (2a) are obvious. We prove that (1a) ⇒ (1c) and (2a) ⇒ (2c).

It is easy to see that, for every polynomially length-bounded function $f : \Sigma^* \to \mathbb{N}$, there exist a set $B \subseteq \Sigma^* \times \Sigma^*$ and a strictly increasing polynomial $p$ such that $f(x) = \|\{z \mid |z| = p(|x|) \wedge (x, z) \in B\}\|$. Note that we may choose $B$ so that, for all $x \in \Sigma^*$, $(x, 0^{p(|x|)}) \notin B$ and $(x, 1^{p(|x|)}) \notin B$. If, in addition, $f \in \#\mathrm{P}$, then $B$ can be chosen from P.

We construct a total p-order $A$ on $\Sigma^*$ as follows. Generally, $A$ will coincide with the lexicographical order on $\Sigma^*$ except that, for every $x \in \Sigma^*$, the interval between $x0^{p(|x|)}$ and $x1^{p(|x|)}$ (inclusively) is ordered differently in the following way.

- First comes $x1^{p(|x|)}$.
- Next come the elements of $\{xz \mid |z| = p(|x|) \wedge (x, z) \in B\}$ in lexicographical order.
- Finally come the elements of $\{xz \mid |z| = p(|x|) \wedge (x, z) \notin B \wedge z \neq 1^{p(|x|)}\}$ in lexicographical order.

Note that $f(x) = \|\{w \mid x1^{p(|x|)} <_A w <_A x0^{p(|x|)}\}\|$. If, in addition, $B \in \mathrm{P}$, then $A \in \mathrm{P}$. ❑

# 5 Polynomial-Time Orders with Efficient Adjacency Checks

From Theorem 4.1, we know that counting the size of intervals with respect to P-decidable partial p-orders that have polynomial-time computable boundaries computes some function in #P. The situation changes if in addition we require each P-decidable partial p-order to have efficient adjacency checks.

**Definition 5.1** $\mathrm{IF_p}$ *(respectively,* $\mathrm{IF_t}$*) is the class of all functions* $f : \Sigma^* \to \mathbb{N}$ *for which there exist a partial (respectively, total) p-order* $A \in \mathrm{P}$ *having efficient adjacency checks and functions* $b, t \in \mathrm{FP}$*, such that, for every* $x \in \Sigma^*$*,* $f(x) = \|\{z \mid b(x) <_A z <_A t(x)\}\|$*.*

The following theorem places the classes $\mathrm{IF_t}$ and $\mathrm{IF_p}$ between two well-known complexity classes.

**Theorem 5.2** $\mathrm{FP} \subseteq \mathrm{IF_t} \subseteq \mathrm{IF_p} \subseteq \#\mathrm{P}$.

*Proof.* The second inclusion follows from the definitions of $\mathrm{IF_t}$ and $\mathrm{IF_p}$, and the third inclusion follows from Theorem 4.1. Thus, it remains to prove that $\mathrm{FP} \subseteq \mathrm{IF_t}$. For each $f \in \mathrm{FP}$, there exists a strictly increasing polynomial $p$ such that $f(x) < 2^{p(|x|)} - 1$. For $x \in \Sigma^*$ and $i < 2^{p(|x|)}$, let $\mathrm{bin}(x, i)$ be the binary description of $i$ having exactly $p(|x|)$ bits.

We construct a total p-order $A$ on $\Sigma^*$ as follows. Generally, $A$ coincides with the lexicographical order on $\Sigma^*$ except that, for every $x \in \Sigma^*$, the interval between $x0^{p(|x|)}$ and $x1^{p(|x|)}$ (inclusively) is ordered in the following way.

- First come the elements of $\{x\mathrm{bin}(x, i) \mid 0 \leq i \leq f(x)\}$ in lexicographical order.
- Next comes $x1^{p(|x|)}$.
- Finally come the elements of $\{x\mathrm{bin}(x, i) \mid f(x) < i < 2^{p(|x|)} - 1\}$ in lexicographical order.

Note that $A$ is P-decidable, has efficient adjacency checks, and satisfies $f(x) = \|\{w \mid x0^{p(|x|)} <_A w <_A x1^{p(|x|)}\}\|$. ❑

What else can we say about the relationships between FP, $\mathrm{IF_t}$, $\mathrm{IF_p}$, and #P? We start by providing a characterization of $\mathrm{IF_p}$ based on an important subset of #P. Let $\mathrm{supp}(f)$ denote the support of $f$, i.e., $\mathrm{supp}(f) = \{x \mid f(x) \neq 0\}$.

**Theorem 5.3** $\mathrm{IF_p} = \{f \in \#\mathrm{P} \mid \mathrm{supp}(f) \in \mathrm{P}\}$.

*Proof.* Suppose that $f \in \mathrm{IF_p}$, via p-order $A \in \mathrm{P}$ having polynomial-time adjacency checks and boundary functions $b, t \in \mathrm{FP}$. Note that $\overline{\mathrm{supp}(f)} = \{x \mid b(x) \prec_A t(x) \vee b(x) \not\leq_A t(x)\}$. Thus, since $A \in \mathrm{P}$ and $A_\prec \in \mathrm{P}$, it follows that $\overline{\mathrm{supp}(f)} \in \mathrm{P}$ and thus that $\mathrm{supp}(f) \in \mathrm{P}$. By Theorem 5.2, $f \in \#\mathrm{P}$. Therefore $\mathrm{IF_p} \subseteq \{f \in \#\mathrm{P} \mid \mathrm{supp}(f) \in \mathrm{P}\}$.

We now show that $\{f \in \#\mathrm{P} \mid \mathrm{supp}(f) \in \mathrm{P}\} \subseteq \mathrm{IF_p}$. Suppose $f \in \#\mathrm{P}$ and $\mathrm{supp}(f) \in \mathrm{P}$. Since $f \in \#\mathrm{P}$, there exists a set $B \subseteq \Sigma^* \times \Sigma^*$ from P and a strictly increasing polynomial $p$ such that $f(x) = \|\{z \mid |z| = p(|x|) \wedge (x, z) \in B\}\|$.

We construct a partial p-order $A$ on $\Sigma^*$ as follows. Generally, $A$ coincides with the lexicographical order on $\Sigma^*$ except that, for every $x \in \Sigma^*$, the interval between $x0^{p(|x|)}00$ and $x1^{p(|x|)}11$ (inclusively) is ordered according to the following rules.

- $x0^{p(|x|)}00 <_A x0^{p(|x|)}01 <_A x0^{p(|x|)}11$.
- The elements from $\{xz10 \mid |z| = p(|x|) \wedge (x, z) \in B\}$ are pairwise incomparable, and all are between $x0^{p(|x|)}01$ and $x0^{p(|x|)}11$.
- The elements from $\{xz10 \mid |z| = p(|x|) \wedge (x, z) \notin B\} \cup \{xz\sigma \mid |z| = p(|x|) \wedge z \neq 0^{p(|x|)} \wedge \sigma \in \{00, 01, 11\}\}$ are pairwise incomparable, and all are between $x0^{p(|x|)}00$ and $x0^{p(|x|)}01$.

Note that $A$ is P-decidable and satisfies $f(x) = \|\{w \mid x0^{p(|x|)}01 <_A w <_A x0^{p(|x|)}11\}\|$. Define $b(x) =_{\mathrm{def}} x0^{p(|x|)}01$ and $t(x) =_{\mathrm{def}} x0^{p(|x|)}11$. For each $x$, we have by the construction of $A$ that $b(x) \prec_A t(x)$ if and only if $f(x) = 0$. Since by assumption $\{x \mid f(x) > 0\} \in \mathrm{P}$ the set $\{x \mid b(x) \prec_A t(x)\}$ belongs to P. By our construction, all other adjacency questions are very easily answered by the obvious, efficient test. So $A_\prec \in \mathrm{P}$. ❑

From this it follows that $\mathrm{IF_p}$ and #P coincide on Nonzero, defined as the set $\{f \mid (\forall x \in \Sigma^*)[f(x) > 0]\}$.

**Corollary 5.4** $\mathrm{IF_p} \cap \mathrm{Nonzero} = \#\mathrm{P} \cap \mathrm{Nonzero}$.

In what follows, we will sometimes write 1 for the function class consisting of precisely the constant function $\lambda x.1$, and we will sometimes write $\mathcal{O}(1)$ for the function class consisting of precisely the functions $\lambda x.0$, $\lambda x.1$, $\lambda x.2$, ....

**Corollary 5.5** *1.* $\#\mathrm{P} \subseteq \mathrm{IF_p} \mathbin{\text{-}} 1$.

*2.* $\#\mathrm{P} \mathbin{\text{-}} \mathcal{O}(1) = \mathrm{IF_p} \mathbin{\text{-}} \mathcal{O}(1)$.

From Theorem 5.2 and Corollary 5.5 we can conclude that $\mathrm{IF_p} \subseteq \mathrm{IF_p} \mathbin{\text{-}} 1$, which is equivalent to saying that $\mathrm{IF_p}$ is closed under increment, i.e., for every $f \in \mathrm{IF_p}$, the function $f'$ is also in $\mathrm{IF_p}$, where, for all $x \in \Sigma^*$, $f'(x) =_{\mathrm{def}} f(x) + 1$.

**Corollary 5.6** *The class* $\mathrm{IF_p}$ *is closed under increment.*

Regarding $\mathrm{IF_t}$, we have the following theorem. Note that this theorem's second part says that the three function classes $\mathrm{IF_t}$, $\mathrm{IF_p}$, and #P are so closely related that in the presence of easy-to-compute subtractive postcomputation adjustments they become the same. Though it is not concerned with interval functions, we commend to the attention of the interested reader a beautiful paper by Ogihara et al. [OTTW96] that studies whether for #P postcomputation adjustments can annihilate even the effects of various operators.

**Theorem 5.7** *1.* $\#\mathrm{P} \subseteq \mathrm{IF_t}$ - FP.

*2.* $\mathrm{IF_t}$ - FP = $\mathrm{IF_p}$ - FP = #P - FP.

*Proof.* (1) For $f : \Sigma^* \to \mathbb{N}$ in #P, there exist a set $B \subseteq \Sigma^* \times \Sigma^*$ from P and a strictly increasing polynomial $p$ such that $f(x) = \|\{z \mid |z| = p(|x|) \wedge (x,z) \in B\}\|$.

We construct a total p-order $A$ on $\Sigma^*$ as follows. Generally, $A$ coincides with the lexicographical order on $\Sigma^*$ except that, for every $x$, the interval between $x0^{p(|x|)+2}$ and $x1^{p(|x|)+2}$ (inclusively) is ordered differently in the following way.

- First come the elements of $\{xz00 \mid |z| = p(|x|)\}$ in lexicographical order.
- Next come the elements of $\{xz11 \mid |z| = p(|x|) \wedge (x,z) \in B\} \cup \{xz01 \mid |z| = p(|x|)\}$ in lexicographical order.
- Finally come the elements of $\{xz11 \mid |z| = p(|x|) \wedge (x,z) \notin B\} \cup \{xz10 \mid |z| = p(|x|)\}$ in lexicographical order.

Note that $A$ is in P, has efficient adjacency checks, and satisfies $\|\{w \mid x1^{p(|x|)}00 <_A w <_A x0^{p(|x|)}10\}\| = f(x) + 2^{p(|x|)}$.

(2) This follows from Theorem 5.2 and part 1 of the present theorem. ❑

**Corollary 5.8** $\mathrm{FP}^{\mathrm{IF_t}} = \mathrm{FP}^{\mathrm{IF_p}} = \mathrm{FP}^{\#\mathrm{P}}$.

The previous results indicate that the computational power of $\mathrm{IF_p}$ and $\mathrm{IF_t}$ are not far from the computational power of #P. Nonetheless, Theorem 5.10 shows that these classes cannot coincide unless P = NP. In the proof of Theorem 5.10 we will draw on the following lemma regarding the application of the $\exists$ operator to $\mathrm{IF_p}$ and $\mathrm{IF_t}$. Comparing Lemma 5.9 with Corollary 5.4 and taking into account that $\exists\cdot\#\mathrm{P} = \mathrm{NP}$, it turns out that it is precisely the possibility that $f(x) = 0$ that makes the classes #P and $\mathrm{IF_p}$ potentially differ.

**Lemma 5.9** $\exists\cdot\mathrm{IF_p} = \exists\cdot\mathrm{IF_t} = \mathrm{P}$.

*Proof.* For $L \in \exists\cdot\mathrm{IF_p}$ there exist a p-order $A \in \mathrm{P}$ having efficient adjacency checks and $b, t \in \mathrm{FP}$ such that, for all $x$, $x \in L \Leftrightarrow \|\{z \mid b(x) <_A z <_A t(x)\}\| > 0$. Thus, for all $x \in \Sigma^*$, $x \in L \Leftrightarrow [b(x) \leq_A t(x)$ and $b(x) \not\prec_A t(x)]$, so $x \in L$ can be checked in polynomial time.

Choose $L \in \mathrm{P}$. Thus $\chi_L \in \mathrm{FP}$. By Theorem 5.2, $\chi_L \in \mathrm{IF_t}$, thus $L \in \exists\cdot\mathrm{IF_t}$. ❑

**Theorem 5.10** *The following statements are equivalent.*

*1.* P = NP.

*2.* $\mathrm{IF_p}$ = #P.

*3.* $\mathrm{IF_t}$ = #P.

*4. Every* P*-decidable partial p-order has efficient adjacency checks.*

*5. Every* P*-decidable total p-order has efficient adjacency checks.*

*Proof.* (1) ⇒ (4) follows from Corollary 3.2.1. (4) ⇒ (5) is immediate from the definitions. (5) ⇒ (3) follows from Theorem 4.1.1. (3) ⇒ (2) follows from Theorem 5.2. To see that (2) ⇒ (1), if $\mathrm{IF_p}$ = #P then $\exists\cdot\mathrm{IF_p} = \exists\cdot\#\mathrm{P}$. By Lemma 5.9 and Proposition 2.2.3 we have P = NP. ❑

We know from Theorem 5.2 that $\mathrm{FP} \subseteq \mathrm{IF_t}$. However, if $\mathrm{IF_t} \subseteq \mathrm{FP}$ or even $\mathrm{IF_t} \subseteq \mathrm{UPSV_t}$, then severe consequences follow.

**Theorem 5.11** *1.* $\mathrm{FP} = \mathrm{IF_t}$ *if and only if* $\mathrm{P} = \mathrm{PP}$.

*2.* $\mathrm{IF_t} \subseteq \mathrm{UPSV_t}$ *if and only if* $\mathrm{UP} = \mathrm{PP}$.

*3.* $\mathrm{UPSV_t} \subseteq \mathrm{IF_p}$ *if and only if* $\mathrm{P} = \mathrm{UP} \cap \mathrm{coUP}$.

*Proof.* For items (1) and (2) we consider the left-to-right direction first. From Theorem 5.7 and Proposition 2.2, we can conclude under the assumption $\mathrm{FP} = \mathrm{IF_t}$ that $\mathrm{PP} = \exists\cdot(\#\mathrm{P} - \mathrm{FP}) = \exists\cdot(\mathrm{IF_t} - \mathrm{FP}) = \exists\cdot(\mathrm{FP} - \mathrm{FP}) = \mathrm{P}$ and we can conclude under the assumption $\mathrm{IF_p} \subseteq \mathrm{UPSV_t}$ that $\mathrm{PP} = \exists\cdot(\#\mathrm{P} - \mathrm{FP}) = \exists\cdot(\mathrm{IF_t} - \mathrm{FP}) \subseteq \exists\cdot(\mathrm{UPSV}_t - \mathrm{FP}) = \mathrm{UP} \cap \mathrm{coUP}$. For the right-to-left directions, if $\mathrm{P} = \mathrm{PP}$, then $\mathrm{IF_t} \subseteq \#\mathrm{P} \subseteq \mathrm{FP}^{\#\mathrm{P}} = \mathrm{FP}^{\mathrm{PP}} = \mathrm{FP}$. Thus, $\mathrm{IF_t} = \mathrm{FP}$. If $\mathrm{UP} = \mathrm{PP}$, then $\mathrm{IF_t} \subseteq \#\mathrm{P} \subseteq \mathrm{FP}^{\#\mathrm{P}} = \mathrm{FP}^{\mathrm{PP}} = \mathrm{FP}^{\mathrm{UP}\cap\mathrm{coUP}} = \mathrm{UPSV_t}$.

For item (3), from $\mathrm{UPSV}_t \subseteq \mathrm{IF_p}$, Proposition 2.2, and Lemma 5.9 it follows that $\mathrm{UP}\cap\mathrm{coUP} = \exists\cdot\mathrm{UPSV}_t \subseteq \exists\cdot\mathrm{IF_p} = \mathrm{P}$. For the right-to-left direction, by Proposition 2.1.1, $\mathrm{P} = \mathrm{UP} \cap \mathrm{coUP}$ implies $\mathrm{UPSV_t} = \mathrm{FP}$. So, by Theorem 5.2, $\mathrm{P} = \mathrm{UP} \cap \mathrm{coUP}$ implies $\mathrm{UPSV_t} \subseteq \mathrm{IF_p}$ (and even $\mathrm{UPSV_t} \subseteq \mathrm{IF_t}$). ❑

In contrast to Theorem 5.11.3, when restricted to strictly positive functions the class $\mathrm{UPSV_t}$ is even included in $\mathrm{IF_t}$.

**Theorem 5.12** $\mathrm{UPSV_t} \cap \mathrm{Nonzero} \subseteq \mathrm{IF_t} \cap \mathrm{Nonzero}$.

*Proof.* Choose $f$ in $\mathrm{UPSV_t}\cap\mathrm{Nonzero}$ and let $M$ be a nondeterministic polynomial-time Turing machine that, for every $x \in \Sigma^*$, produces an output on exactly one computation path, and this output is $f(x)$. Without loss of generality, suppose that all computation paths of $M$ on input $x \in \Sigma^*$ have length exactly $p(|x|)$, where $p$ is a strictly increasing polynomial. For $x \in \Sigma^*$ and $i < 2^{p(|x|)}$, let $\mathrm{bin}(x, i)$ be the $p(|x|)$-bit binary description of $i$. Observe that the set $B =_{\mathrm{def}} \{xz\mathrm{bin}(x, i) \mid |z| = p(|x|)$ and $M$ on input $x$ produces along computation path $z$ an output and that output is lexicographically strictly greater than $i\}$ is in P and that $f(x) = \|\{y \mid |y| = 2p(|x|) \wedge xy \in B\}\|$.

We construct a total p-order $A$ on $\Sigma^*$ as follows. Generally, $A$ coincides with the lexicographical order on $\Sigma^*$ except that, for every $x \in \Sigma^*$, the interval between $x0^{2p(|x|)+2}$ and $x1^{2p(|x|)+2}$ (inclusively) is ordered differently in the following way.

- First come the elements of $\{xzu00 \mid |z| = |u| = p(|x|)\}$ in lexicographical order.
- Next come the elements of $\{xzu11 \mid |z| = |u| = p(|x|) \wedge xzu \in B\}$ in lexicographical order.
- Next come the elements of $\{xzu01 \mid |z| = |u| = p(|x|)\}$ in lexicographical order.
- Next come the elements of $\{xzu11 \mid |z| = |u| = p(|x|) \wedge xzu \notin B\}$ in lexicographical order.
- Finally come the elements of $\{xzu10 \mid |z| = |u| = p(|x|)\}$ in lexicographical order.

Note that $A$ is in P, has efficient adjacency checks, and satisfies $\|\{w \mid x1^{2p(|x|)}00 <_A w <_A x0^{2p(|x|)}01\}\| = \|\{y \mid |y| = 2p(|x|) \wedge xy \in B\}\| = f(x)$. ❑

Since $\mathrm{UPSV_t}$ is closed under increment, Theorem 5.12 yields the following corollary.

**Corollary 5.13** $\mathrm{UPSV_t} \subseteq \mathrm{IF_t} \mathbin{\text{-}} 1$.

Corollary 5.6 showed that the class $\mathrm{IF_p}$ is closed under increment. This is also true for the class $\mathrm{IF_t}$.

**Theorem 5.14** *The class* $\mathrm{IF_t}$ *is closed under increment.*

*Proof.* For $f \in \mathrm{IF_t}$ there exist a P-decidable p-order $A$ on $\Sigma^*$ with efficient adjacency checks and functions $b, t \in \mathrm{FP}$ such that, for all $x \in \Sigma^*$, $f(x) = \|\{w \mid b(x) <_A w <_A t(x)\}\|$. Without loss of generality we may require that $b(x) \leq_A t(x)$, since on inputs not satisfying that we may modify $t(x)$ to output $b(x)$. Let $p$ be a strictly increasing polynomial such that, for all $y \in \Sigma^*$ satisfying $y \leq_A t(x)$, $|y| < p(|x|)$.

We construct a total p-order $A'$ on $\Sigma^*$ as follows. Generally, $A'$ coincides with the lexicographical order on $\Sigma^*$ except that, for every $x \in \Sigma^*$, the interval between $x0^{p(|x|)+2}$ and $x1^{p(|x|)+2}$ (inclusively) is ordered in the following way.

- First comes $x0^{p(|x|)+2}$.
- Next come the elements of $D_x =_{\mathrm{def}} \{x0^{p(|x|)-|z|}1z0 \mid b(x) \leq_A z \leq_A t(x)\}$, for which we set $x0^{p(|x|)-|y|}1y0 \leq_{A'} x0^{p(|x|)-|z|}1z0$ if and only if $y \leq_A z$.
- Finally come the elements of $\{xu \mid |u| = p(|x|)+2\} - (D_x \cup \{0^{p(|x|)+2}\})$ in lexicographical order.

Note that $A'$ is P-decidable, has efficient adjacency checks, and that $f(x)+1 = \|\{w \mid x0^{p(|x|)+2} <_{A'} w <_{A'} x0^{p(|x|)-|t(x)|}1t(x)0\}\|$. ❑

**Corollary 5.15** $\mathrm{IF_t} \subseteq \mathrm{IF_t} \mathbin{\text{-}} 1$.

Although the statement "$\mathrm{UPSV_t} = \mathrm{IF_t}$" is not likely to be true (see Theorem 5.11), for the case of strictly positive, polynomially bounded functions the analogous statement holds. We define $\mathrm{PolyBounded} =_{\mathrm{def}} \{f \mid (\exists \text{ polynomial } p)(\forall x)[f(x) \leq p(|x|)]\}$.

**Theorem 5.16** *1.* $\mathrm{IF_t} \cap \mathrm{PolyBounded} \subseteq \mathrm{UPSV_t} \cap \mathrm{PolyBounded}$.

*2.* $\mathrm{IF_t} \cap \mathrm{PolyBounded} \cap \mathrm{Nonzero} = \mathrm{UPSV_t} \cap \mathrm{PolyBounded} \cap \mathrm{Nonzero}$.

*3.* $\mathrm{UPSV_t} \cap \mathrm{PolyBounded} \subseteq \mathrm{IF_p} \cap \mathrm{PolyBounded}$ *if and only if* $\mathrm{P} = \mathrm{UP} \cap \mathrm{coUP}$.

*Proof.* For item (1), let $f$ be a polynomially bounded function, i.e., there is a polynomial $p$ such that, for all $x \in \Sigma^*$, $f(x) \leq p(|x|)$, and let $f$ be in $\mathrm{IF_t}$ via total p-order $A \in \mathrm{P}$ having efficient adjacency checks, and functions $b, t \in \mathrm{FP}$. Let $q$ be a polynomial such that, for all $x$ and $y$, $(x, y) \in A$ implies $|x| \leq q(|y|)$. Define $M$ to be a machine that, on input $x$, does the following.

(a) Nondeterministically guess an integer $m$ such that $m \leq p(|x|)$,

(b) if $m = 0 \wedge (t(x) \leq_A b(x) \vee b(x) \prec_A t(x))$, then accept and output 0.

(c) if $m > 0$ then nondeterministically guess $m$ distinct strings $z_1, \ldots, z_m$ with $|z_i| \leq q(|t(x)|)$, and check whether $b(x) \prec_A z_1 \prec_A z_2 \prec_A \cdots \prec_A z_m \prec_A t(x)$, and if so accept and output $m$.

(d) Reject.

Since $A \in \mathrm{P}$ and $A$ has efficient adjacency checks, $M$ runs in nondeterministic polynomial time, and since $A$ is a total p-order there exists, with respect to $\prec_A$, at most one chain between $b(x)$ and $t(x)$. So one can see that $M$ on input $x \in \Sigma^*$ has exactly one accepting path and the output on the path is precisely $f(x)$. Thus, $f \in \mathrm{UPSV_t}$.

(2): The inclusion "$\supseteq$" follows from Theorem 5.12, and the inclusion "$\subseteq$" follows from part 1 of the present theorem.

(3): For the "only if" direction, let $L$ be a $\mathrm{UP} \cap \mathrm{coUP}$ set. Then its characteristic function $\chi_L$ is trivially polynomially bounded and is in $\mathrm{UPSV}_t$, and so is, by the assumption, in $\mathrm{IF_p}$. Thus, there are a P-decidable partial p-order $A$ having efficient adjacency checks and polynomial-time computable functions $b, t$ such that $c_L(x) = \|\{z \mid b(x) <_A z <_A t(x)\}\|$. Consequently, $\chi_L(x) = 1 \Leftrightarrow (b(x) \leq_A t(x) \wedge b(x) \not\prec_A t(x))$. The "if" direction follows from the "if" direction Theorem 5.11.3. ❑

From Theorem 4.1 we know that total p-orders that are efficiently decidable and partial p-orders that are efficiently decidable describe the same class of functions in our setting (namely #P). If we consider p-orders that additionally have efficient adjacency checks, then the analogous confluence of total and partial does not hold unless an unexpected complexity class collapse occurs.

**Theorem 5.17** *If* $\mathrm{IF_t} = \mathrm{IF_p}$*, then* $\mathrm{UP} = \mathrm{PH}$.

*Proof.* Assume that $\mathrm{IF_t} = \mathrm{IF_p}$. We show that $\mathrm{coNP} \subseteq \mathrm{UP}$ (which is equivalent to the statement $\mathrm{UP} = \mathrm{PH}$). Let $L \in \mathrm{coNP}$, i.e., there is a function $f \in \#\mathrm{P}$ such that, for all $x \in \Sigma^*$, $x \in L \Leftrightarrow f(x) = 0$. Consider the function $f'$, where $f'(x) =_{\mathrm{def}} f(x) + 1$. Thus $x \in L \Leftrightarrow f'(x) = 1$ and, since #P is closed under increment, we conclude that $f' \in \#\mathrm{P} \cap \mathrm{Nonzero} = \mathrm{IF_p} \cap \mathrm{Nonzero} = \mathrm{IF_t} \cap \mathrm{Nonzero}$. Thus, there exist a total p-order $A \in \mathrm{P}$ with efficient adjacency checks and functions $b, t \in \mathrm{FP}$ such that $f'(x) = \|\{z \mid b(x) <_A z <_A t(x)\}\|$. Let $q$ be a polynomial such that $(x, y) \in A$ implies $|x| \leq q(|y|)$. Define $M$ to be a machine that, on input $x \in \Sigma^*$, nondeterministically guesses $z$ such that $|z| \leq q(|t(x)|)$ and checks whether $b(x) \prec_A z \prec_A t(x)$. Clearly, $M$ runs in polynomial time (since $A$ has efficient adjacency checks) and always has at most one accepting path (since $A$ is a total p-ordering and we are doing two adjacency checks in our test). Moreover, $x \in L$ if and only if $M$ on $x$ has an accepting computation path. Thus, $L \in \mathrm{UP}$. ❑

# 6 Arbitrary Orders with Efficient Adjacency Checks

In the previous section, we studied polynomial-time-decidable p-orders having efficient adjacency checks. We showed that the classes defined by interval size functions over such orders, $\mathrm{IF_p}$ and $\mathrm{IF_t}$, are very close to #P. In the present section, we consider what happens when we do not insist on polynomial-time decidability for the order but still require efficient adjacency checks. Section 6.1 presents our results on this. Due to its complexity and length, the proof of one key claim of that section, Lemma 6.5, is presented separately as Section 6.2.

## 6.1 Results on Arbitrary Orders with Efficient Adjacency Checks

In this section, we study p-orders that have efficient adjacency checks, but that are not required to be polynomial-time decidable. We define two classes to capture this behavior.

**Definition 6.1** *The class* $\mathrm{IF}_\mathrm{p}^*$ *(respectively,* $\mathrm{IF}_\mathrm{t}^*$*) is the set of all functions* $f : \Sigma^* \to \mathbb{N}$ *for which there exist a partial (respectively, total) p-order* $A$ *having efficient adjacency checks and functions* $b, t \in \mathrm{FP}$ *such that, for every* $x \in \Sigma^*$, $f(x) = \|\{z \mid b(x) <_A z <_A t(x)\}\|$.

We have the following inclusions between classes of interval size functions and other complexity classes of functions.

**Proposition 6.2** $\mathrm{IF_t} \subseteq \mathrm{IF}_\mathrm{t}^* \subseteq \mathrm{IF}_\mathrm{p}^* \subseteq \mathrm{FPSPACE(poly)}$ *and* $\mathrm{IF_t} \subseteq \mathrm{IF_p} \subseteq \mathrm{IF}_\mathrm{p}^* \cap \#\mathrm{P} \subseteq \#\mathrm{P} \subseteq \mathrm{FPSPACE(poly)}$.

*Proof.* The only inclusion that is nontrivial is $\mathrm{IF}_\mathrm{p}^* \subseteq \mathrm{FPSPACE(poly)}$. Let $f$ be in $\mathrm{IF}_\mathrm{p}^*$ via a partial p-order $A$ having efficient adjacency checks and functions $b, t \in \mathrm{FP}$. Let $p$ be a polynomial such that, for all $x, y \in \Sigma^*$, $(x, y) \in A$ implies $|x| \leq p(|y|)$. From Proposition 3.1 we know that $A$ is in PSPACE. Thus, there is a polynomial-space Turing machine $M$ that, for any input $x \in \Sigma^*$, counts by brute force how many strings $z$ of length at most $p(|t(x)|)$ satisfy $b(x) <_A z <_A t(x)$. We may thus conclude that $f$ is in FPSPACE(poly). ❑

The main results of this section show that the computational powers of $\mathrm{IF}_\mathrm{p}^*$ and $\mathrm{IF}_\mathrm{t}^*$ are close to the computational power of FPSPACE(poly). In fact, within the flexibility of the simple post-computation adjustment of subtracting polynomial-time computable functions, these three classes become the same.

**Theorem 6.3** $\mathrm{IF}_\mathrm{t}^* - \mathrm{FP} = \mathrm{IF}_\mathrm{p}^* - \mathrm{FP} = \mathrm{FPSPACE(poly)} - \mathrm{FP}$.

**Theorem 6.4** $\exists \cdot \mathrm{IF}_\mathrm{t}^* = \exists \cdot \mathrm{IF}_\mathrm{p}^* = \mathrm{PSPACE}$.

Theorems 6.3 and 6.4 follow immediately from Proposition 6.2 and the following lemma, whose proof is deferred to Section 6.2.

**Lemma 6.5** *For each* $f \in \mathrm{FPSPACE(poly)}$*, there exist a total p-order* $A$ *having efficient adjacency checks and polynomial-time computable functions* $s : \mathbb{N} \to \mathbb{N}$, $b : \Sigma^* \to \Sigma^*$, $b' : \Sigma^* \to \Sigma^*$, *and* $t : \Sigma^* \to \Sigma^*$ *such that, for all* $x \in \Sigma^*$,

1. *$s$ is polynomially bounded.*
2. $\|\{z \mid b(x) <_A z <_A t(x)\}\| = 2^{2s(|x|)+1} + f(x) - 2$, *and*
3. $\|\{z \mid b'(x) <_A z <_A t(x)\}\| > 0$ *if and only if* $f(x) = 1$.

As a consequence of Theorems 6.3 and 6.4, we obtain characterizations for the class FPSPACE(poly) in terms of $\mathrm{IF}_\mathrm{t}^*$. For classes $\mathcal{F}$ and $\mathcal{G}$ of functions from $\Sigma^*$ to $\mathbb{N}$, let $\mathcal{F} \ominus \mathcal{G}$ denote the class of all total, nonnegative functions in $\mathcal{F} - \mathcal{G}$, i.e., the class of all total functions $h$ for which there exist total functions $f \in \mathcal{F}$ and $g \in \mathcal{G}$ such that, for all $x \in \Sigma^*$, $f(x) \geq g(x)$ and $h(x) = f(x) - g(x)$.

**Corollary 6.6**

1. $\mathrm{FPSPACE(poly)} = \mathrm{IF}_\mathrm{t}^* \ominus \mathrm{FP} = \mathrm{FP}^{\mathrm{IF}_\mathrm{t}^*} = \mathrm{FP}^{\exists \cdot \mathrm{IF}_\mathrm{t}^*}$.
2. $\mathrm{FPSPACE(poly)} = \mathrm{IF}_\mathrm{p}^* \ominus \mathrm{FP} = \mathrm{FP}^{\mathrm{IF}_\mathrm{p}^*} = \mathrm{FP}^{\exists \cdot \mathrm{IF}_\mathrm{p}^*}$.

*Proof.* Regarding part 1, by Theorem 6.3, Proposition 6.2, and Theorem 6.4 we have $\mathrm{FPSPACE(poly)} \subseteq \mathrm{IF}_\mathrm{t}^* \ominus \mathrm{FP} \subseteq \mathrm{FP}^{\mathrm{IF}_\mathrm{t}^*} \subseteq \mathrm{FP}^{\mathrm{FPSPACE(poly)}} \subseteq \mathrm{FP}^{\mathrm{PSPACE}} \subseteq \mathrm{FP}^{\exists \cdot \mathrm{IF}_\mathrm{t}^*} \subseteq \mathrm{FP}^{\mathrm{PSPACE}} \subseteq \mathrm{FPSPACE(poly)}$. Part 2 holds by the same inclusion chain applied to $\mathrm{IF}_\mathrm{p}^*$. ❑

Though Theorem 6.3 shows that $\mathrm{IF}_t^*$ is almost as powerful as FPSPACE(poly), the following theorem shows that it is unlikely that $\mathrm{IF}_t^*$ actually coincides with FPSPACE(poly).

**Theorem 6.7** *If* FPSPACE(poly) $\subseteq \mathrm{IF}_p^*$ *then* UP = PSPACE.

*Proof.* Suppose that FPSPACE(poly) $\subseteq \mathrm{IF}_p^*$. Let $L \in$ PSPACE. Then its characteristic function $\chi_L$ is in FPSPACE(poly), and by hypothesis $\chi_L \in \mathrm{IF}_p^*$ via some partial p-order $A$ having efficient adjacency checks, some polynomial $p$ such that $(x, y) \in A$ implies $|x| \leq p(|y|)$, and functions $b, t \in$ FP such that $\chi_L(x) = \|\{z \mid b(x) <_A z <_A t(x)\}\|$. Note that $L = \{x \mid (\exists z)[|z| \leq p(|t(x)|) \wedge b(x) \prec_A z \prec_A t(x)]\}$. Thus, keeping in mind that $(\forall x)[\chi_L(x) \leq 1]$, we have $L \in$ UP. ❑

From Theorems 6.3 and 6.4, if $\mathrm{IF}_t^* = \mathrm{IF}_t$ or $\mathrm{IF}_t^* \subseteq \#\mathrm{P} - \mathrm{FP}$, then strong consequences follow, as the following two corollaries show.

**Corollary 6.8** *The following statements are equivalent.*

1. P = PSPACE.
2. $\mathrm{IF}_p = \mathrm{IF}_p^*$.
3. $\mathrm{IF}_t = \mathrm{IF}_t^*$.
4. *Every partial p-order with efficient adjacency checks is* P*-decidable.*
5. *Every total p-order with efficient adjacency checks is* P*-decidable.*

*Proof.* (1) ⇒ (4) is just Corollary 3.2.2. (4) ⇒ (5) is trivial. (4) ⇒ (2) and (5) ⇒ (3) follow from the definitions of $\mathrm{IF}_p$, $\mathrm{IF}_p^*$, $\mathrm{IF}_t$, and $\mathrm{IF}_t^*$. By Theorem 6.4 and Lemma 5.9, (2) implies PSPACE $= \exists \cdot \mathrm{IF}_p^* = \exists \cdot \mathrm{IF}_p =$ P and so implies (1). Similarly, (3) implies PSPACE $= \exists \cdot \mathrm{IF}_t^* = \exists \cdot \mathrm{IF}_t =$ P and so implies (1). ❑

**Corollary 6.9** *1. If* $\mathrm{IF}_t^* \subseteq \#\mathrm{P} - \mathrm{FP}$*, then* SPP = PSPACE*.*

*2. If* $\mathrm{IF}_t^* \subseteq \#\mathrm{P}$*, then* NP = SPP = PSPACE*.*

*Proof.* (1): For $L \in$ PSPACE, $\chi_L \in$ FPSPACE(poly). By Proposition 6.2, Theorem 6.3, and our assumption, $\chi_L \in \#\mathrm{P} - \mathrm{FP}$. Thus, $L \in$ SPP.

(2): From Theorem 6.4 and our hypothesis, we obtain PSPACE $\subseteq \exists \cdot \#\mathrm{P} =$ NP. Combining this with the first part of this theorem we have SPP = NP = PSPACE. ❑

The next result is analogous to results regarding the potential equality of $\mathrm{IF}_t$ and $\mathrm{IF}_p$.

**Theorem 6.10** *If* $\mathrm{IF}_t^* = \mathrm{IF}_p^*$*, then* UP = PH*.*

*Proof.* The proof follows the proof of Theorem 5.17, except that, for the function there called $f'$, we now conclude that $f' \in \#\mathrm{P} \cap \mathrm{Nonzero} = \mathrm{IF}_p \cap \mathrm{Nonzero} \subseteq \mathrm{IF}_p^* \cap \mathrm{Nonzero} = \mathrm{IF}_t^* \cap \mathrm{Nonzero}$. This approach works because the hypothesis $f' \in \mathrm{IF}_t^*$ can be exploited in the same way as the hypothesis $f' \in \mathrm{IF}_t$ was exploited in the proof of Theorem 5.17. This is because in the proof of Theorem 5.17 the P-decidability of the total p-order underlying $f' \in \mathrm{IF}_t$ was not even used. ❑

Figure 1 summarizes the results we have obtained regarding the inclusion structure of our classes. Although we have not proven consequences of collapses other than those drawn in the figure, we conjecture that the inclusions in the figure are all one can prove without assuming unexpected collapses of complexity classes.

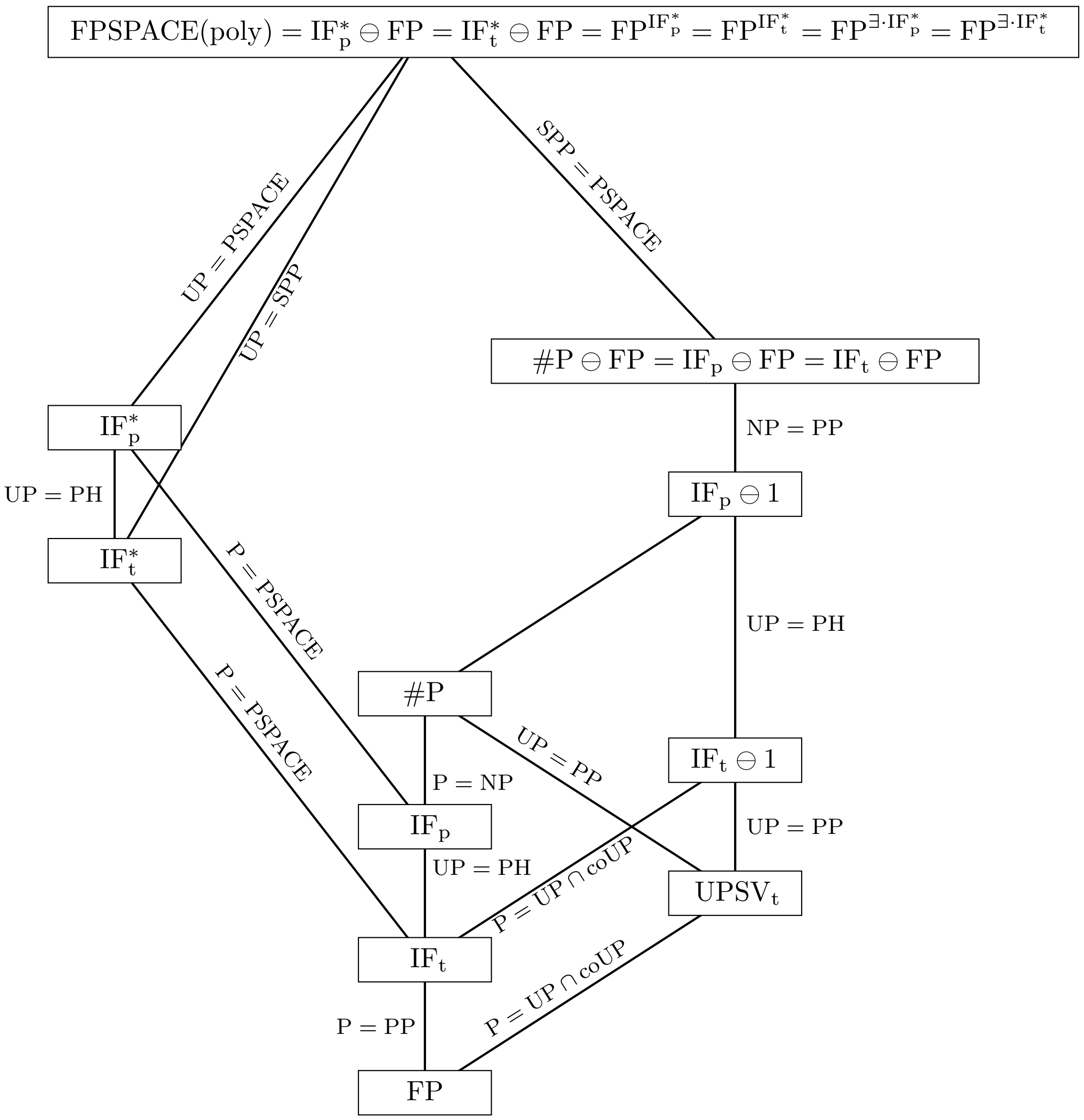

Figure 1: The landscape of interval size function classes and related function classes. An equation $E$ on the edge between the function classes $\mathcal{F}_1$ and $\mathcal{F}_2$ means that $\mathcal{F}_1 = \mathcal{F}_2$ implies $E$. The edge equations that are not immediate consequences of the results of this paper are well-known or easy to see. Since FP, which forms the base of this containment tower, is of type $\Sigma^* \to \mathbb{N}$, the fact that in the above figure we use "$\ominus$" rather than "$-$" is of no consequence.

## 6.2 Proof of Lemma 6.5

The goal of this section is to prove Lemma 6.5. For convenience, we repeat its statement here.

**Lemma 6.5** For each $f \in$ FPSPACE(poly), there exist a total p-order $A$ having efficient adjacency checks and polynomial-time computable functions $s : \mathbb{N} \to \mathbb{N}$, $b : \Sigma^* \to \Sigma^*$, $b' : \Sigma^* \to \Sigma^*$, and $t : \Sigma^* \to \Sigma^*$ such that, for all $x \in \Sigma^*$,

1. $s$ is polynomially bounded.
2. $\|\{z \mid b(x) <_A z <_A t(x)\}\| = 2^{2s(|x|)+1} + f(x) - 2$, and
3. $|\{z \mid b'(x) <_A z <_A t(x)\}\| > 0$ if and only if $f(x) = 1$.

Constructing the p-order $A$ mentioned in Lemma 6.5 is, compared to the other p-orders described in this paper, more technically involved. Before we prove Lemma 6.5, we will show, for any $f \in$ FPSPACE(poly), how to construct $A$ based on the behavior of a Turing machine that computes $f$. We will then prove Lemma 6.5 by showing that $A$ has all the properties claimed by the lemma.

We will construct $A$ in five phases, described as follows.

1. **Fixing the Computational Model**. We will base $A$ on a Turing machine $M$ that computes $f$ in a natural but somewhat nonstandard way. The benefit of using $M$ rather than an arbitrary FPSPACE(poly) Turing machine for $f$ is that it will be easier to work with binary encodings of the configurations of $M$ and the actions of $M$ than with those of an arbitrary FPSPACE(poly) Turing machine for $f$.

2. **Fixing the Encoding**. We will base $A$ on binary encodings of the configurations of $M$, which we call *enhanced instantaneous descriptions*. Our encodings are like standard instantaneous descriptions (IDs) [HMU01] but differ in three crucial ways. First, our encodings are actual binary strings rather than sequences of abstract symbols. Second, we use different syntax (which we describe below). Finally, our descriptions contain more information than is actually needed to describe a configuration of $M$ at an instant in time. This additional information is never accessed by $M$, so its presence in the encodings does not affect the performance of $M$. At the same time, its presence will greatly aid us in constructing $A$.

3. **Building Trees**. For some appropriate polynomial $s$, we will, for each $x \in \Sigma^*$, define a tree whose nodes are enhanced instantaneous descriptions of $M$ and whose edges are based on the next move function of $M$. This tree will have a subtree $T_x$ having exactly $2^{2s(|x|)}$ nodes.

4. **Traversing the Trees**. We will associate multiple strings with each node in the tree described above (by padding the labels of the nodes) in such a way that $f(|x|) + 2$ strings are associated with one of the nodes in $T_x$ and two strings are associated with each of the remaining $2^{2s(|x|)} - 1$ nodes in $T_x$. We will then define a total, one-to-one, polynomial-time computable function $D_M$ over these strings in such a way that $D_M$, applied repeatedly to some appropriate starting point, represents a traversal of the tree such that the traversal visits each of these strings once, i.e., from a particular one of the strings $z$ associated with the root of the tree, for each string $y$ associated with some node of the tree there is an integer $i \in \mathbb{N}$ such that $D_M^{(i)}(z) = y$, where $D_M^{(0)}(z) = z$, and, for each $i \in \{1, 2, 3, \ldots\}$, $D_M^{(i)}(z) = D_M(D_M^{(i-1)}(z))$. Moreover, for strings $w$ and $y$, $D_M(w) = y$ only if the nodes associated with $w$ and $y$ are related (i.e., parent/child, sibling, or identical nodes).

5. **Constructing $\boldsymbol{A}$**. We will base $A$ on $D_M$. For example, $A$ crucially will have the property that if $w$ and $z$ are two of the strings described in Phase 4, then $w \prec_A z$ if and only if $z = D_M(w)$. Note there will also be many strings on which $D_M$ is not defined that will nonetheless have to be accounted for. Through careful encoding at each phase in the construction, it will be easy to account for these strings in such a way that $A$ has all the properties we desire.

After we handle these five phases, we will prove Lemma 6.5. We now proceed with the construction. Please note that, due to the length of this construction, we overload certain variables. For instance, the variable $t$ denotes both a function over strings and over natural numbers, and has distinct semantics in each case. Over strings it is the function that determines the "bottom" of an interval (i.e., it is used as it typically is throughout this paper), and over the natural numbers it bounds the amount of space needed for part of the encodings we use.

## Phase 1: Fixing the Computational Model

Let $M = (Q, \Sigma, \Gamma, \delta, B, q_0, F)$ be a Turing machine that computes $f$, where

- $Q$ is the set of *state symbols*,
- $\Sigma = \{0, 1\}$ is the set of *input symbols*,
- $B$ is the *blank symbol*,
- $\Gamma \supseteq \{0, 1, B\}$ is the set of allowable *tape symbols*,
- $\delta$ is the *next move function*, i.e., a mapping from $Q \times \Gamma$ to $Q \times \Gamma \times \{-1, 1\}$,
- $q_0$ is the *start state*, and
- $F \subseteq Q$ is the set of *final states*.

We assume that $M$ has the following properties.

- For some $m \in \mathbb{N}$, $||Q|| = ||\Gamma|| = 2^m$ (any Turing machine not having this property can be turned into one having this property by adding extra "dummy" states and symbols to its current sets of state and tape symbols, respectively). Since $\Gamma \supseteq \{0, 1, B\}$, $m \geq 2$.
- $F$ contains a single element, $q_f$, and $q_0 \neq q_f$.
- $M$ has a single, one-way infinite tape (a standard PSPACE(poly) Turing machine would have distinct input, output, and work tapes). On no input $x$ does a true run of $M$ move off the left end of the tape. (One way to ensure that $M$ has this latter property is to include the symbols, $0^e$, $1^e$, and $B^e$ in $\Gamma$. These symbols will be used, exactly on the leftmost cell of the tape, as replacements for 0, 1, and $B$. We can then construct $M$ so that it is in its start state just once, namely at the beginning of the run, and that, from its start state, it always replaces the then-current symbol (which, in a true run, will always be located in the leftmost tape cell and will be either 0, 1, or $B$) not with whatever symbol it would normally write during that step but rather with the appropriate analog among $0^e$, $1^e$, and $B^e$. Similarly, our machines can be forced to be such that they attempt to ensure that at all future times this left-marking is preserved, i.e., a $0^e/1^e/B^e$-marker square may be changed during the run but just among $0^e$, $1^e$, and $B^e$, as appropriate. A Turing machine constructed in this way can, on any true run, determine when it is about to (were it to mindlessly perform the simulation of the underlying machine) move off the left end, and can indeed handle—without itself running off the left end and in a fashion that is consistent in effect with whatever standard behavior (typically either rejection or "bouncing off" the left end) we in our notion of Turing machines associate with attempting to go off the left end—the left-end move-off that was about to happen.

- $\delta$ on input $(q, r) \in Q \times \Gamma$ is defined if and only if $(q, r) \notin \{q_f\} \times \Gamma$.
- For all $r \in \Gamma$ and all $i \in \{-1, 1\}$, $(q_0, r, i)$ is not in the image of $\delta$. (That is, nothing moves *to* the start state.)
- For all $x \in \Sigma^*$, $M$ on input $x$ halts with $y \in \Sigma^*$ written on its $|y|$ leftmost tape cells, where $y$ is the shortest binary representation of $f(x)$ (i.e., no leading zeros, unless $f(x) = 0$), and with every other tape cell containing the blank symbol.
- There is a strictly increasing polynomial $p$ such that, on each input $x \in \Sigma^*$, $M$ uses, at most, $p(|x|)$ tape cells and $p(|x|) > 0$.

## Phase 2: Fixing the Encoding

We now describe the binary encoding we use to describe the configurations of $M$. Figure 2 provides an overview of this phase of the construction. Let $\varphi : Q \to \{0,1\}^m$ be a total bijection (recall that $||Q|| = 2^m$ and $m \geq 2$) such that $\varphi(q_0) = 0^m$ and $\varphi(q_f) = 1^m$. The function $\varphi^{-1}$ denotes the unique total bijection from $\{0,1\}^m$ to $Q$ that inverts $\varphi$. Let $\theta : \Gamma \to \{0,1\}^m$ be a total bijection (recall that $||\Gamma|| = 2^m$ and $m \geq 2$) such that $\theta(B) = 0^m$, $\theta(0) = 1^{m-1}0$, and $\theta(1) = 1^m$. Define $\hat{\theta} : \Gamma^* \to (\{0,1\}^m)^*$ recursively as $\hat{\theta}(\epsilon) = \epsilon$, and, for all $y \in \Gamma$ and $w \in \Gamma^*$, $\hat{\theta}(wy) = \hat{\theta}(w)\theta(y)$. Since $\hat{\theta}$ is also a bijection, we use $\hat{\theta}^{-1}$ to denote the unique total bijection from $(\{0,1\}^m)^*$ to $\Gamma^*$ that inverts $\hat{\theta}$.

We define the "partially encoded" next move function $\delta' : \{0,1\}^m \times \{0,1\}^m \to \{0,1\}^m \times \{0,1\}^m \times \{-1,1\}$ on input $(q,r) \in \{0,1\}^m \times \{0,1\}^m$ as $\delta'(q,r) = (\varphi(q'), \theta(r'), i)$, where $q'$, $r'$, and $i$ are specified by $\delta(\varphi^{-1}(q), \theta^{-1}(r)) = (q', r', i)$.

Recall that $\Sigma = \{0,1\}$. Define $\nu : \Gamma^* \to \mathbb{N}$ recursively as $\nu(\epsilon) = 0$ and, for each $y \in \Gamma$ and $w \in \Gamma^*$,

$$\nu(wy) = \begin{cases} 1 + 2\nu(w) & \text{if } y = 1 \wedge w \in \Sigma^* \\ 2\nu(w) & \text{if } y = 0 \wedge w \in \Sigma^* \\ \nu(w) & \text{if } y = B \\ 0 & \text{otherwise.} \end{cases}$$

This has the property that if $z \in \Sigma^* B^*$, then $\nu(z)$ is the natural number that $z$ represents in binary. And if $z \in \Gamma^* - \Sigma^* B^*$, then $\nu(z) = 0$.

We also need the following notation. For any domain $S$, any (possibly partial) function $h : S \to S$, any $i \in \mathbb{N}$, and any $s \in S$, we define $h^{(i)}(s)$ as

$$h^{(i)}(s) \quad =_{\text{def}} \quad \begin{cases} s & \text{if } i = 0 \\ h(h^{(i-1)}(s)) & \text{if } i > 0 \wedge (h^{(i-1)}(s) \text{ is defined}) \wedge (h^{(i-1)}(s) \in \text{domain}(h)) \\ \text{undefined} & \text{otherwise.} \end{cases}$$

Note that if $h(a)$ is undefined then so, for example, will be $h^{(1)}(a)$ and $h^{(2)}(a)$.

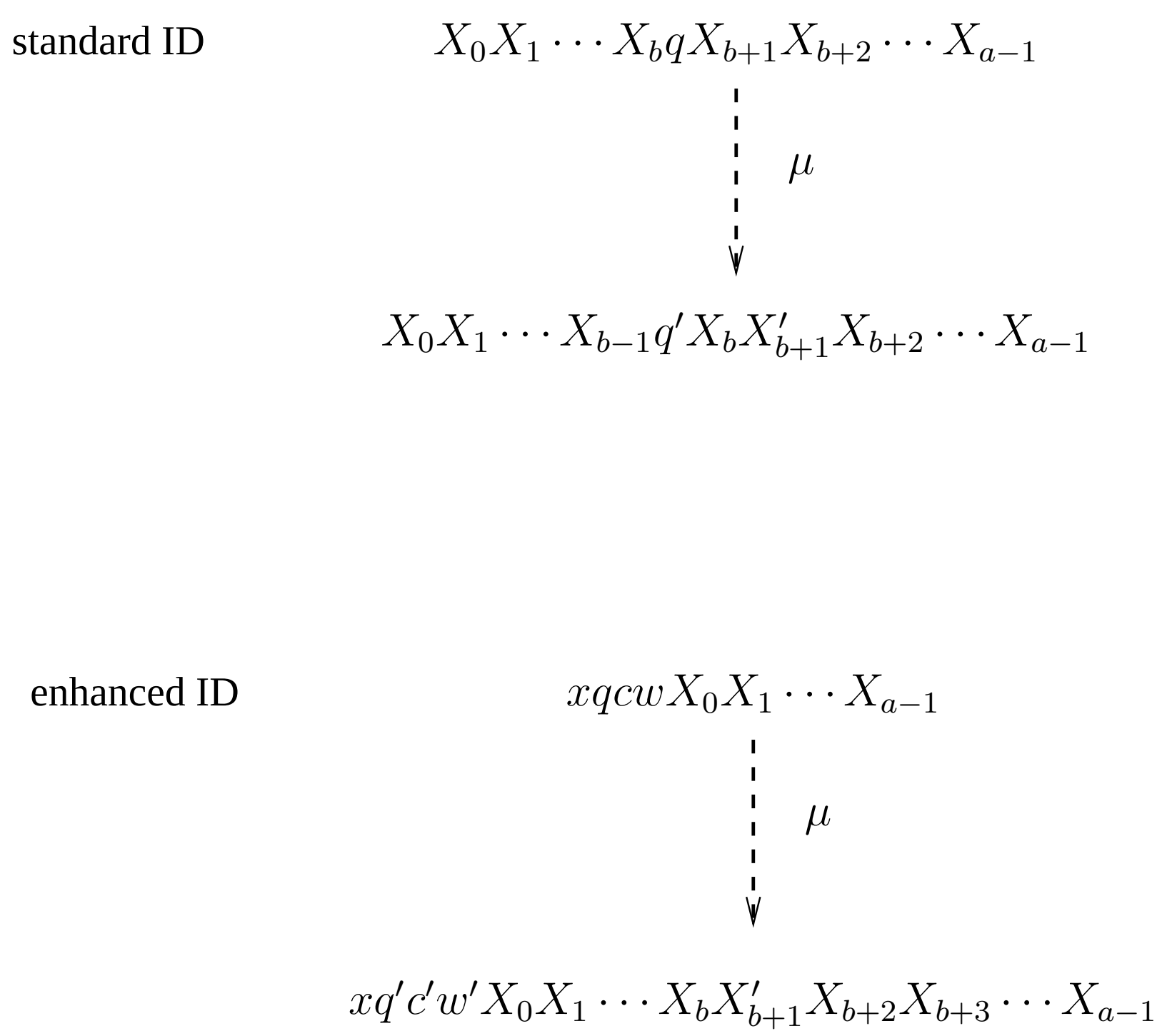

Figure 2: A brief comparison between standard instantaneous descriptions (IDs) and the enhanced IDs we use. Before the computation step illustrated, the tape head is at cell $b+1$ and the machine is in state $q$. Afterwards, the head is at cell $b$ and the machine is in state $q'$. The symbol $\mu$ represents the next move function. In standard IDs, the state $q$ appears immediately before the tape cell that the head is currently visiting (e.g., in the case illustrated above, cell $b+1$ before the move and $b$ afterwards). Our enhanced IDs contain additional strings: $x$, $c$, and $w$. The string $x$ encodes the input to the Turing machine, $c$ encodes the number of computation steps the Turing machine has performed so far, and $w$ is the position of the tape head. The state string remains in the same place throughout the computation, and instead $w$ is updated with the position of the tape head. Thus, $w$ encodes the number $b+1$ (i.e., the position of the tape head before the computation step), and $w'$ encodes $b$ (i.e., the position of the tape head after the computation step). The strings $c$ and $c'$ also represent numbers, where the number encoded by $c'$ is one greater than the number encoded by $c$. For more details on eIDs and encodings, see the text.

All logarithms in this paper are base two, i.e., $\log m$ means $\log_2 m$. Define functions $r$, $s$, and $t$ on input $n \in \mathbb{N}$ as $r(n) =_{\mathrm{def}} \lceil \log p(n) \rceil$ (recall that, by assumption, on any input of length $n$, $M$ uses at most $p(n)$ tape cells and $p(n) > 0$), $t(n) =_{\mathrm{def}} m2^{r(n)}$, and $s(n) =_{\mathrm{def}} m + r(n) + t(n)$.

Let $\mathbb{eID} =_{\mathrm{def}} \bigcup_{n=0}^{\infty} \{0,1\}^{n+2s(n)}$ be the set of *enhanced instantaneous descriptions* of $M$. Informally speaking, for each $n \in \mathbb{N}$ and $x \in \Sigma^n$, $q \in \{0,1\}^m$, $c \in \Sigma^{s(n)}$, $w \in \Sigma^{r(n)}$, and $X_0, X_1, \ldots, X_{2^{r(n)}-1} \in \Sigma^m$, the string $xqcwX_0X_1 \cdots X_{2^{r(n)}-1} \in \mathbb{eID}$ is interpreted as follows.

- The string $x$ represents the input to $f$.
- The string $q$ represents the instantaneous state of $M$.
- The string $c$ will be used as an external clock ("external" because it is not maintained by $M$ itself, but rather by an "outside observer") to count the number of computational steps $M$ has made so far. The presence of the external clock will allow us to adapt the next move function of $M$ to the enhanced instantaneous descriptions of $M$ in such a way that cycles never occur, even if $M$ from a particular configuration may cycle. Note that, since the number of tape cells $M$ uses is polynomially bounded in the length of its input, we only need a polynomial amount of bits for the clock. Intuitively speaking, if the clock "runs out of time" by running out of bits, then (assuming we chose a large enough polynomial to control the number of clock bits) we know that a cycle has occurred.
- The string $w$ encodes the instantaneous position of the tape head, i.e., a position of 0 or 1 or ... or $2^{r(|x|)} - 1$ is encoded (respectively) by the string $0^{r(x)}$ or $0^{r(x)-1}1$ or ... or $1^{r(x)}$.
- The strings $X_0, X_1, \ldots, X_{2^{r(n)}-1}$ represent the instantaneous contents of the leftmost $2^{r(n)}$ tape cells of $M$.

Note that the second, fourth, and fifth sections of the string described above (i.e., $q$, $w$, and $X_0, X_1, \ldots, X_{2^{r(n)}-1}$) are already sufficient to describe $M$ at any instant. Note also that, because $s$, $r$, and $t$ are all polynomial-time computable and nondecreasing, we can, in polynomial time, for each $n \in \mathbb{N}$ and each $z \in \Sigma^{n+2s(n)}$, compute from $z$ the value $n$ and the locations of the five above-described sections of $z$, and these locations are well-defined.

For each $x \in \Sigma^*$, we call $x0^m0^{s(|x|)}0^{r(|x|)}\varphi(x)0^{t(|x|)-|\varphi(x)|} = x0^{2s(|x|)-t(|x|)}\varphi(x)0^{t(|x|)-|\varphi(x)|} \in \mathbb{eID}$ the *initial configuration* of $M$ on $x$, denoted $i_{M,x}$. The string $i_{M,x}$ represents a configuration on which $M$ would be started under "normal usage." Note that $\mathbb{eID}$ contains strings that represent configurations of $M$ that are never reached under "normal usage." From these "unreachable" configurations, $M$ may run forever or attempt to move off the left end of the tape. (Note that the true run of $M$ on input $x$ certainly does not run forever, since $M$ is computing an FPSPACE(poly) function and FPSPACE(poly) is a class of total functions, and our model of function computing requires $M$ to halt in order for it to compute a value. Recall that we assume that on no true run of $M$ on input $x$ will $M$ attempt to move off the left end of the tape. We did not explicitly discuss the semantics of attempting to move off the left end of the tape, but the point of the comment above is that even if our model of computing FPSPACE(poly) functions is such that moving off the left end of the tape is considered like running forever and makes a function be undefined on the input, and so never happens on a true run of a machine computing an FPSPACE(poly) function, it nonetheless may be the case that such a machine when started at some "unreachable" configuration might attempt to run off the left end of the tape.)

We define a *move* over $\mathbb{eID}$ via a function $\mu : \Sigma^* \to \Sigma^*$ that we will define now. An important consideration in the design of $\mu$ is to exploit the additional information present in the enhanced IDs to guarantee that $\mu$ never loops and that it always "ends" (i.e., returns the value undefined) "gracefully" (in an sense that will soon become clear, including, for example, that it does not blindly try to move off the left end of the tape).

For each $x \in \Sigma^*$, $c \in \{0,1\}^{s(|x|)}$, $w \in \{0,1\}^{r(|x|)}$, $X_0, X_1, \ldots, X_{2^{r(|x|)}-1} \in \{0,1\}^m$, and $q \in \{0,1\}^m - \{1^m\}$,

$$\mu(xqcwX_0X_1 \cdots X_{2^{r(|x|)}-1}) =_{\mathrm{def}} xq'c'w'X_0X_1 \cdots X_{\nu(w)-1}YX_{\nu(w)+1}X_{\nu(w)+2} \cdots X_{2^{r(|x|)}-1} \quad (1)$$

if

$$\delta'(q, X_{\nu(w)}) \text{ is defined } \wedge c \neq 1^{s(|x|)} \wedge 0 \leq \nu(w) + i < 2^{r(|x|)}, \quad (2)$$

where

$$\delta'(q, X_{\nu(w)}) = (q', Y, i),\ c' \in \{0,1\}^{s(|x|)},\ w' \in \{0,1\}^{r(|x|)},\ \nu(c') = \nu(c) + 1, \text{ and } \nu(w') = \nu(w) + i,$$

and

$$\mu(xqcwX_0X_1 \cdots X_{2^{r(|x|)}-1}) =_{\mathrm{def}} x1^mcwX_0X_1 \cdots X_{2^{r(|x|)}-1} \quad (3)$$

otherwise. If $q = 1^m$, $\mu(xqcwX_0X_1 \cdots X_{2^{r(|x|)}-1})$ is undefined. For all $y \notin \mathbb{eID}$, $\mu(y)$ is undefined. It is easy to see that the behavior of $\mu$ described by equation 1 is roughly analogous to the behavior of $\delta$. Indeed, for all $x \in \Sigma^*$, there exists a number $j \in \mathbb{N}$ such that $\mu^{(j)}(i_{M,x}) = x1^mcwz$, where $c \in \{0,1\}^{s(|x|)}$, $w \in \{0,1\}^{r(|x|)}$, $z \in \{0,1\}^{t(|x|)}$, $\nu(c) = j$, and $\nu(\hat{\theta}^{-1}(z)) = f(x)$. Equation 3 enforces "gracefulness" by detecting when the configuration encoded by the input string is about to move off the left end of the tape or is about to use too much tape or has a "$c$" value that has already reached $2^{s(|x|)}$ (note that no actual run can ever run more than $2^{s(n)}$ steps without running forever, but running forever can never happen on actual runs since all functions in FPSPACE(poly) are total). In such cases, $\mu$ simply changes the state bits to represent the final state (i.e., $1^m$).

Proposition 6.11 collects several easy-to-see properties of $\mu$.

**Proposition 6.11** *1. The function $\mu$ is polynomial-time computable.*

2. *The function $\mu$ is length-preserving, i.e., for all $w \in \Sigma^*$, if $\mu(w)$ is defined, then $|w| = |\mu(w)|$.*

3. *For all $x \in \Sigma^*$, all $w \in \{0,1\}^{2s(|x|)-m}$, and all $q \in \{0,1\}^m$, $\mu(xqw)$ is defined if and only if $q \neq 1^m$.*

4. *For all $w \in \Sigma^*$, there exists a number $j$ such that $\mu^{(j)}(w)$ is undefined.*

5. *In polynomial time we can, for each $z \in \Sigma^*$, enumerate all $y$ such that $\mu(y) = z$.*

6. *For each $w \in \mathbb{eID}$ and each $j \in \mathbb{N}^+$, if $\mu^{(j)}(w)$ is defined, then $\mu^{(j)}(w) \neq w$.*

*Proof.* All items are easy to see. However, item 5 deserves some additional explanation. To perform this enumeration, if $z \notin \mathbb{eID}$, then there is no $y$ such that $\mu(y) = z$. If $z \in \mathbb{eID}$, then examine the next move function of $M$ to determine the configurations from which $M$ in one step will move into the configuration encoded by $z$. There are only a constant number of such configurations. Output the strings of length $|z|$ that encode these configurations. This takes care of all preimages of $z$ that satisfy equation 2. If, for some $x \in \Sigma^*$, $c \in \{0,1\}^{s(|x|)}$, $w \in \{0,1\}^{r(|x|)}$, and $X_0, X_1, \ldots, X_{2^{r(|x|)}-1} \in \{0,1\}^m$ it holds that $z = x1^mcwX_0X_1 \cdots X_{2^{r(|x|)}-1}$ (i.e., if $z$ satisfies the conditions of equation 3) then, for each $q \in \{0,1\}^m - \{1^m\}$ such that $xqcwX_0X_1 \cdots X_{2^{r(|x|)}-1}$ does not satisfy equation 2, output $xqcwX_0X_1 \cdots X_{2^{r(|x|)}-1}$. This takes care of all preimages of $z$ that do not satisfy equation 2. ❑

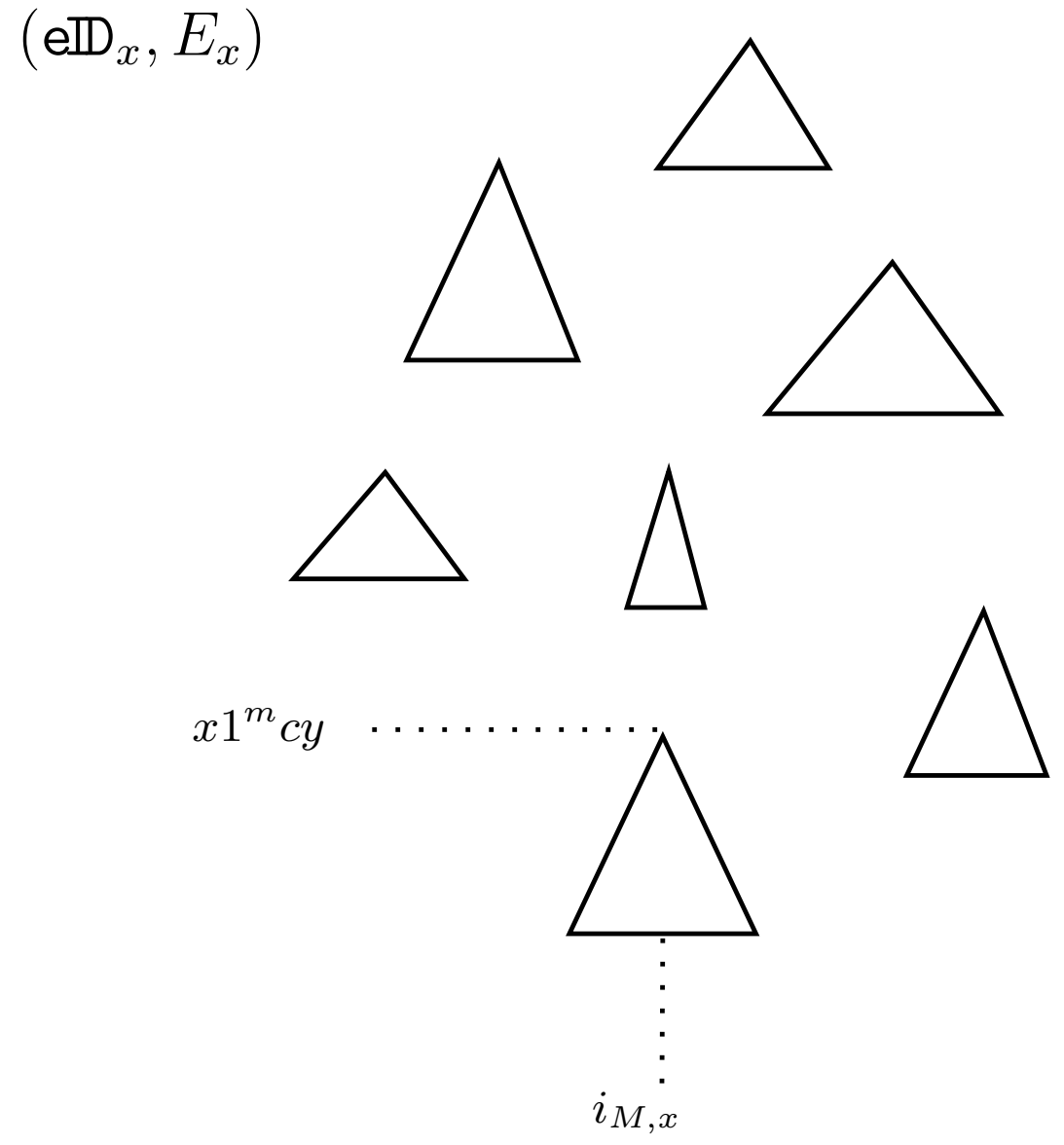

Figure 3: The directed forest $(\mathrm{eID}_x, E_x)$. Note that precisely one tree in the digraph $(\mathrm{eID}_x, E_x)$ has $i_{M,x}$ as a node, and note that in that tree $i_{M,x}$ will be a leaf node. For some $c$ and $y$ satisfying $c \in \{0,1\}^{2s(|x|)-t(|x|)-m}$, $y \in \{0,1\}^{t(|x|)}$, and $\nu(\hat{\theta}^{-1}(y)) = f(x)$, that tree will have as its root node $x1^m cy$.

## Phase 3: Building Trees

For each $x \in \Sigma^*$, let

$$\mathrm{eID}_x = \{xw \mid w \in \{0,1\}^{2s(|x|)}\}$$

and

$$E_x = \{(xw, xz) \mid xw, xz \in \mathrm{eID}_x \wedge \mu(xw) = xz\}.$$

A directed forest is an acyclic digraph in which all nodes have outdegree at most one. Note that the digraph $(\mathrm{eID}_x, E_x)$ has outdegree at most one. By Proposition 6.11.6, $(\mathrm{eID}_x, E_x)$ is acyclic. Thus, $(\mathrm{eID}_x, E_x)$ is a directed forest (see Figure 3).

For each $x \in \Sigma^*$, let (keep in mind that given the string $xw \in \mathrm{eID}$, it is easy to identify $x$ and $w$)

$$\mathrm{eID}'_x = \{xwy \mid xw \in \mathrm{eID}_x \wedge y \in \{0,1\}^{t(|x|)}\}$$

and

$$E'_x = \{(xwy, xzy) \mid w \in \{0,1\}^{2s(|x|)} \wedge y \in \{0,1\}^{t(|x|)} \wedge xwy \in \mathrm{eID}'_x \wedge \mu(xw) = xz\}.$$

Note that the digraph $(\mathrm{eID}'_x, E'_x)$ is a directed forest, and that, for each tree in $(\mathrm{eID}_x, E_x)$, there are exactly $2^{t(|x|)}$ corresponding trees in $(\mathrm{eID}'_x, E'_x)$ (see Figure 4 for a pictorial preview of this part of the construction).

Let $\mathrm{R}_x =_{\mathrm{def}} \{xwy \in \mathrm{eID}'_x \mid w \in \{0,1\}^{2s(|x|)} \wedge y \in \{0,1\}^{t(|x|)} \wedge (\mu(xw) \text{ is undefined})\}$. Note that, by Proposition 6.11.3, $\mathrm{R}_x = \{xwy \in \mathrm{eID}'_x \mid w \in \{0,1\}^{2s(|x|)} \wedge y \in \{0,1\}^{t(|x|)} \wedge (xw \text{ is the root of a tree in } (\mathrm{eID}_x, E_x))\} = \{x1^m wy \in \mathrm{eID}'_x \mid w \in \{0,1\}^{2s(|x|)-m} \wedge y \in \{0,1\}^{t(|x|)}\}$. Let $\leq_{\mathrm{R}_x}$ denote the order (with $<_{\mathrm{R}_x}$ and $\prec_{\mathrm{R}_x}$ denoting the corresponding "less than" and "predecessor" relations, respectively) defined over $\mathrm{R}_x$ that is determined

by the following sequence. (The reader is cautioned that in what follows "$w$" is used as a variable to catch substrings of various lengths other than the $2s(|x|)$-length strings it has been primarily used for so far).

- First come the elements of $\{xwyy \in \mathrm{R}_x \mid w \in \{0,1\}^{2s(|x|)-t(|x|)} \wedge y \in \{0,1\}^{t(|x|)}\}$ in lexicographic order. Note that the last element in this sequence is $x1^{2s(|x|)+t(|x|)}$.
- Next come the elements of $\{xwdy \in \mathrm{R}_x \mid w \in \{0,1\}^{2s(|x|)-t(|x|)} \wedge d,y \in \{0,1\}^{t(|x|)} \wedge d \neq y\}$ in lexicographic order. Note that the last element in this sequence is $x1^{2s(|x|)+t(|x|)-1}0$.

For each $x \in \Sigma^*$, $w \in \{0,1\}^{2s(|x|)}$, and $y \in \{0,1\}^{t(|x|)}$, we define $\mu_1 : \Sigma^* \to \Sigma^*$, on input $xwy$, as

$$\mu_1(xwy) = \begin{cases} \mu(xw)y & \text{if } xwy \notin \mathrm{R}_x \\ xz & \text{if } xwy \in \mathrm{R}_x \wedge xwy \neq x1^{2s(|x|)+t(|x|)-1}0, \text{ where } xwy \prec_{R_x} xz. \end{cases}$$

In all other cases, $\mu_1$ is undefined. Informally speaking, $\mu_1$ is an "augmented next move" function based on $\mu$, but with the difference that $\mu_1$ in effect strings together all the trees in $(\mathrm{e\mathbb{D}}'_x, E'_x)$ into one giant tree $T_{M,x}$ (see Figure 4 again).

**Proposition 6.12** *For each $x \in \Sigma^*$, let $E''_x =_{\mathrm{def}} \{(w,z) \mid w \in \mathrm{e\mathbb{D}}'_x \wedge \mu_1(w) = z\}$, and define $T_{M,x}$ to be the digraph $(\mathrm{e\mathbb{D}}'_x, E''_x)$.*

1. *The function $\mu_1$ is polynomial-time computable.*
2. *The function $\mu_1$ is length-preserving (i.e., on inputs $a$ for which it is not undefined, $|\mu_1(a)| = |a|$).*
3. *In polynomial time we can, for any $z \in \Sigma^*$, enumerate all $y \in \Sigma^*$ such that $\mu_1(y) = z$.*
4. *For every $x \in \Sigma^*$ and every $w \in \{0,1\}^{2s(|x|)+t(|x|)}$, there exists a number $j \in \mathbb{N}$ such that $\mu_1^{(j)}(xw) = x1^{2s(|x|)+t(|x|)-1}0$. (See also Figure 4.)*
5. *For every $x \in \Sigma^*$ and every $w \in \{0,1\}^{2s(|x|)+t(|x|)}$, $\mu_1(xw)$ is undefined if and only if $w = 1^{2s(|x|)+t(|x|)-1}0$.*
6. *For each $x \in \Sigma^*$ and each $w \in \{0,1\}^{2s(|x|)}$, there is a unique $y \in \{0,1\}^{t(|x|)}$ such that, for some $k \in \mathbb{N}$, $\mu_1^{(k)}(xwy) = x1^{2s(|x|)+t(|x|)}$. (Again, viewing Figure 4—paying particular attention to the black trees—will help make this clear).*
7. *For each $x \in \Sigma^*$, $\|\{w \mid (\exists j \in \mathbb{N})[\mu_1^{(j)}(w) = x1^{2s(|x|)+t(|x|)}]\}\| = 2^{2s(|x|)}$.*
8. *For each $x \in \Sigma^*$, the unique (by item 6) $y \in \{0,1\}^{t(|x|)}$, and each $k \in \mathbb{N}$ such that $\mu_1^{(k)}(i_{M,x}y) = x1^{2s(|x|)+t(|x|)}$, it holds that $f(x) = \nu(\hat{\theta}^{-1}(y))$.*
9. *For each $x \in \Sigma^*$, the digraph $T_{M,x}$ is a tree.*
10. *The subtree of $T_{M,x}$ rooted at $x1^{2s(|x|)+t(|x|)}$ has exactly $2^{2s(|x|)}$ nodes.*

*Proof.* Items 1–5 follow from the definition of $\mu_1$.

For item 6, choose an arbitrary $x \in \Sigma^*$, $w \in \{0,1\}^{2s(|x|)}$, and $y \in \{0,1\}^{t(|x|)}$, and let $j \in \mathbb{N}$, $v \in \{0,1\}^{2s(|x|)-t(|x|)}$, and $d \in \{0,1\}^{t(|x|)}$ be such that $\mu^{(j)}(xw) = xvd$ and $\mu(xvd)$ is undefined (such $j$, $v$, and $d$ exist by Propositions 6.11.4 and 6.11.2). By the definition of $\mathrm{R}_x$, $xvdy \in \mathrm{R}_x$. By the definition of $\leq_{\mathrm{R}_x}$, $\mu^{(j)}(xw)d \leq_{\mathrm{R}_x} x1^{2s(|x|)+t(|x|)}$ and so, by the definition of $\mu_1$, there exists a number $k \geq j$ such that $\mu_1^{(k)}(xwd) = x1^{2s(|x|)+t(|x|)}$. On the other hand, for all $y \in \{0,1\}^{t(|x|)}$ such that $y \neq d$, by the definition of $\leq_{\mathrm{R}_x}$, $x1^{2s(|x|)+t(|x|)} <_{\mathrm{R}_x} \mu^{(j)}(xw)y$, and so, by items 4 and 5 (which guarantee that $\mu_1$ does not cycle), there is no $k$ such that $\mu_1^{(k)}(xwy) = x1^{2s(|x|)+t(|x|)}$.

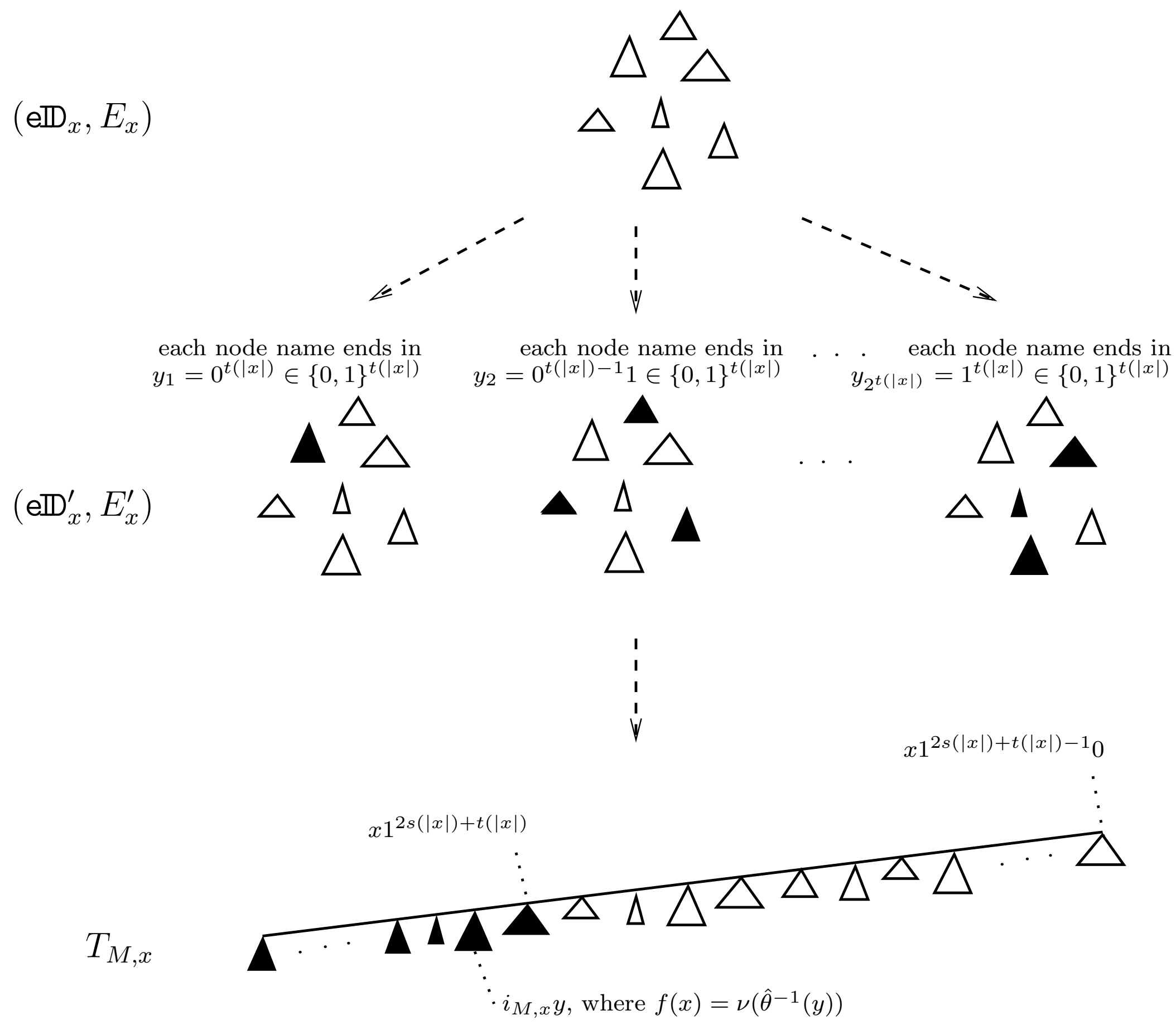

Figure 4: Transforming the directed forest $(\mathrm{e}\mathbb{ID}_x, E_x)$ into $T_{M,x}$. First, $2^{t(|x|)}$ copies of each tree in $(\mathrm{e}\mathbb{ID}_x, E_x)$ are made by appending $t(|x|)$ "guess" bits to each node in each original tree, creating the directed forest $(\mathrm{e}\mathbb{ID}'_x, E'_x)$. Next, the trees in $(\mathrm{e}\mathbb{ID}'_x, E'_x)$ are strung together into a single tree $T_{M,x}$ in such a way that a subtree of $T_{M,x}$ is formed by the trees in $(\mathrm{e}\mathbb{ID}'_x, E'_x)$ having (note: $\mathrm{R}_x$ will be defined in the main text) roots in $\{xwyy \in \mathrm{R}_x \mid w \in \{0,1\}^{2(|s|)-t(|x|)} \wedge y \in \{0,1\}^{t(|x|)}\}$ (represented in the figure by the black trees), i.e., the trees whose "guess" bits equal the contents of the machine tape at the end of the computation. This subtree has exactly one node for each string in $\mathrm{e}\mathbb{ID}_x$, including $i_{M,x}$, and the node associated with $i_{M,x}$ has as its "guess" bits the true output of $M$ on input $x$. We will later exploit this information when we define a traversal of this tree.

Item 7 follows from item 6.

For item 8, choose an arbitrary $x \in \Sigma^*$, and by item 6 let $y$ be the unique member of $\{0,1\}^{t(|x|)}$ such that, for some $k \in \mathbb{N}$, $\mu_1^{(k)}(i_{M,x}y) = x1^{2s(|x|)+t(|x|)}$. Choose $j \in \mathbb{N}$ such that $\mu^{(j)}(i_{M,x})y \in \mathrm{R}_x$. By the definition of $\mu_1$, there exists a number $v \in \{0,1\}^{2s(|x|)-t(|x|)}$ such that $\mu^{(j)}(i_{M,x}) = xvy$ and, by the definition of $\mu$, $M$ on input $x$ halts with $y$ on its tape. Thus, $f(x) = \nu(\hat{\theta}^{-1}(y))$.

Item 9 follows from items 4 and 5.

Item 10 follows from item 7 and the observation that, for any $x \in \Sigma^*$ and any $w, y \in \mathtt{e}\mathbb{ID}'$, $w$ is in the subtree of $T_{M,x}$ rooted at $y$ if and only if $y$ is a node of $T_{M,x}$ and there exists a number $k \in \mathbb{N}$ such that $\mu_1^{(k)}(w) = y$. ❑

## Phase 4: Defining a Traversal

We define $\mathtt{dwn} : \Sigma^* \to \Sigma^* \cup \{\bot\}$, on input $w$, as

$$\mathtt{dwn}(w) = \begin{cases} \max_{\mathrm{lex}} \mu_1^{-1}(w) & \text{if } \mu_1^{-1}(w) \neq \emptyset \\ \bot & \text{otherwise,} \end{cases}$$

where $\max_{\mathrm{lex}}$ returns the maximal element (with respect to the lexicographical order) of a set of strings and we define $\mathtt{acr} : \Sigma^* \to \Sigma^* \cup \{\bot\}$ on input $w$ as

$$\mathtt{acr}(w) = \begin{cases} \max_{\mathrm{lex}}\{w' \mid w' \in \mu_1^{-1}(\mu_1(w)) \wedge w' <_{\mathrm{lex}} w\} & \text{if } \mu_1(w) \text{ is defined} \wedge w \neq \min_{\mathrm{lex}} \mu_1^{-1}(\mu_1(w)) \\ \bot & \text{otherwise,} \end{cases}$$

where $\min_{\mathrm{lex}}$ returns the minimal element (with respect to the lexicographical order) of a set of strings. Clearly, both $\mathtt{dwn}$ and $\mathtt{acr}$ are polynomial-time computable. The function $\mathtt{dwn}$ is named "$\mathtt{dwn}$" because it describes a descent down the tree $T_{M,x}$, and $\mathtt{acr}$ is named "$\mathtt{acr}$" because it describes movement across the tree (i.e., from one sibling node to another). Note that, for all $x \in \Sigma^*$ and all $w \in \{0,1\}^{2s(|x|)+t(|x|)}$ satisfying $xw \in \mathrm{R}_x - \{x1^m0^{2s(|x|)+t(|x|)-m}\}$, it holds that $\mathtt{dwn}(xw) \in \mathrm{R}_x$.

Now, for each $x \in \Sigma^*$, $w \in \{0,1\}^{2s(|x|)}$, $a \in \{0,1\}$, and $y, z \in \{0,1\}^{t(|x|)}$, we define $D_M : \Sigma^* \to \Sigma^*$, a "depth-first"-like traversal of $T_{M,x}$, on input $xwyza$, as

$$D_M(xwyza) = \begin{cases} \mathtt{dwn}(xwy)z0 & \text{if } a = 0 \wedge \mathtt{dwn}(xwy) \neq \bot \wedge \nu(z) = 0 \\ xwyz1 & \text{if } a = 0 \wedge \mathtt{dwn}(xwy) = \bot \wedge xw \neq i_{M,x} \wedge \nu(z) = 0 \\ xwyz'0 & \text{if } a = 0 \wedge \mathtt{dwn}(xwy) = \bot \wedge xw = i_{M,x} \wedge \nu(z) < \nu(\hat{\theta}^{-1}(y)), \text{ where } z \prec_{\mathrm{lex}} z' \\ xwy0^{t(|x|)}1 & \text{if } a = 0 \wedge \mathtt{dwn}(xwy) = \bot \wedge xw = i_{M,x} \wedge \nu(z) = \nu(\hat{\theta}^{-1}(y)) \\ \mathtt{acr}(xwy)z0 & \text{if } a = 1 \wedge \mathtt{acr}(xwy) \neq \bot \wedge \nu(z) = 0 \\ \mu_1(xwy)z1 & \text{if } a = 1 \wedge \mathtt{acr}(xwy) = \bot \wedge \nu(z) = 0. \end{cases}$$

On all other inputs, $D_M$ is undefined.

**Proposition 6.13** *1. The function $D_M$ is polynomial-time computable.*

*2. The function $D_M$ is length-preserving (i.e., for each $v$, either $D_M(v)$ is undefined or $|D_M(v)| = |v|$).*

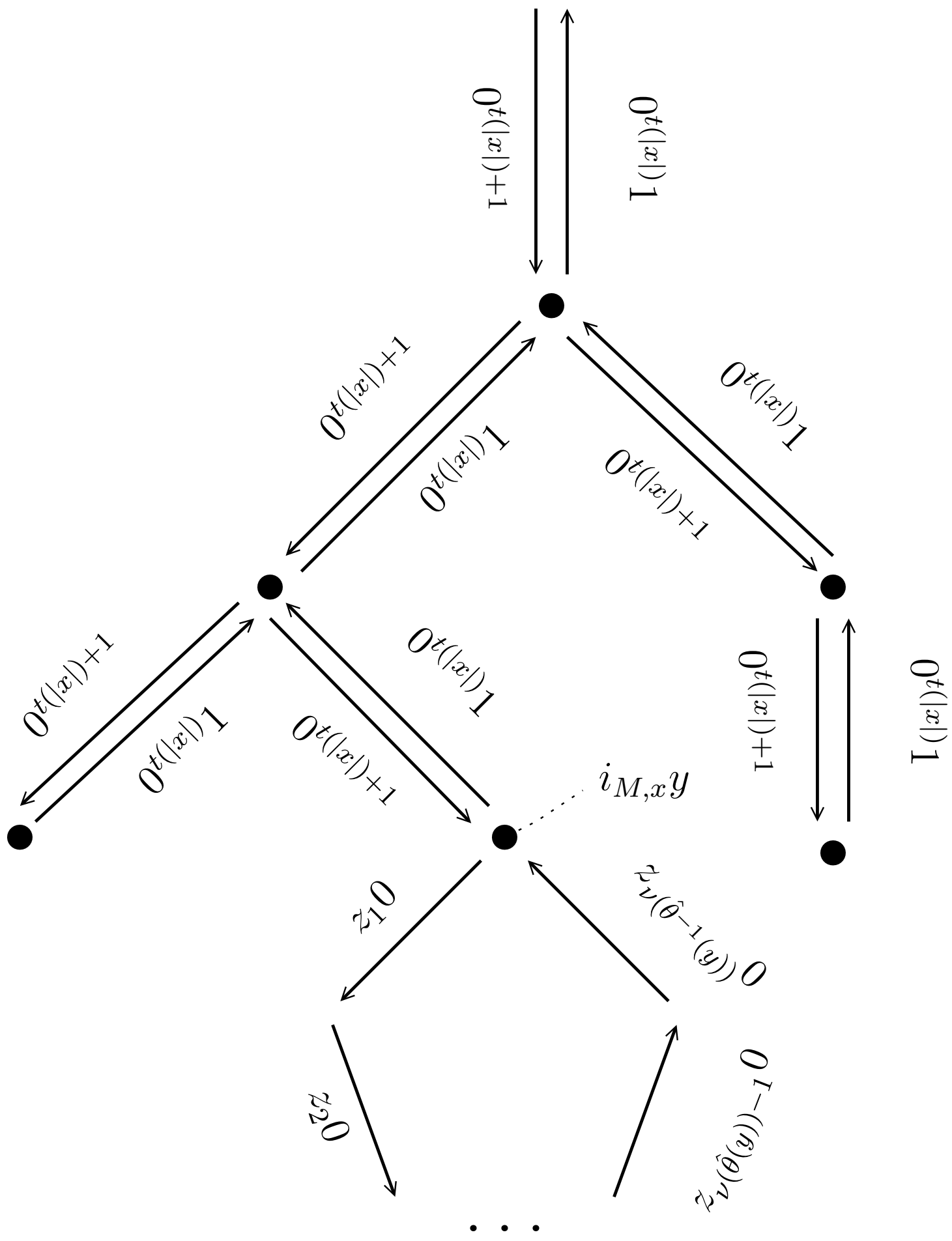

Figure 5: The traversal described by $D_M$. Pictured is a portion of $T_{M,x}$ that contains a node in the initial configuration. The arrows represent the strings associated with the node below them (in the case of the initial configuration node, the arrows below are also associated with it) by padding. The string that is the actual padding appears next to each arrow. $D_M$ is defined over these padded strings. The last bit of each padding string can by seen as controlling the "direction" in which $D_M$ "moves." Note that $y \in \{0,1\}^{t(|x|)}$ and $z_1 = 0^{t(|x|)-1}1, z_2 = 0^{t(|x|)-2}10, \ldots$.

3. *For each $x \in \Sigma^*$, each subtree (and here we really mean each subtree, i.e., not just those corresponding to the trees in digraph $(\mathbb{eID}'_x, E'_x)$—the purpose of this item is to provide insight into how $D_M$ describes a traversal of $T_{M,x}$) $T$ of $T_{M,x}$, each $w \in \{0,1\}^{2s(|x|)}$, and each $y \in \{0,1\}^{t(|x|)}$, $xwy$ is a node of $T$ if and only if there exist $i$, $j$, and $k$ such that $|v| = |xwy|$ (where $v$ is the root of $T$), $D_M^{(i)}(v0^{t(|x|)+1}) = xwy0^{t(|x|)+1}$, $D_M^{(j)}(xwy0^{t(|x|)+1}) = xwy0^{t(|x|)}1$, and $D_M^{(k)}(xwy0^{t(|x|)}1) = v0^{t(|x|)}1$.*

4. *For every $x \in \Sigma^*$, $w \in \{0,1\}^{2s(|x|)}$, $a \in \{0,1\}$, and $y, z \in \{0,1\}^{t(|x|)}$, $D_M(xwyza)$ is defined if and only if $xwyza = x1^{2s(|x|)+t(|x|)-1}0^{t(|x|)+2} \vee (wy \in \{0,1\}^{2s(|x|)+t(|x|)} - \{1^{2s(|x|)+t(|x|)-1}0\} \wedge z = 0^{t(|x|)}) \vee (xw = i_{M,x} \wedge \nu(z) \leq \nu(\hat{\theta}^{-1}(y)) \wedge a = 0)$.*

5. *For every $x \in \Sigma^*$ and every $w \in \{0,1\}^{2(s(|x|)+t(|x|))+1}$, if $D_M(xw)$ is defined, then there exists an $i \in \mathbb{N}$ such that $D_M^{(i)}(x1^{2s(|x|)+t(|x|)-1}0^{t(|x|)+2}) = xw$.*

6. *For all $x \in \Sigma^*$, all $w \in \{0,1\}^{2s(|x|)+t(|x|)}$, all $z \in \{0,1\}^{t(|x|)+1}$, and all $i \in \mathbb{N}$, if $D_M^{(i)}(xwz) = x1^{2s(|x|)+t(|x|)}0^{t(|x|)+1}$, then $xw \in \mathrm{R}_x$.*

7. *The function $\lambda y. \min_{\mathrm{lex}}\{w \mid y <_{\mathrm{lex}} w \wedge (D_M(w)$ is undefined$)\}$ is polynomial-time computable.*

*Proof.* Items 1 and 2 follow from the definition of $D_M$.

For item 3, choose an arbitrary $x \in \Sigma^*$. We prove item 3 by induction over the depth of the subtrees of $T_{M,x}$.

For the base case, choose an arbitrary subtree $T$ of $T_{M,x}$ having depth 1. Let $v$ be the (only) node of $T$. Thus, $\mathtt{dwn}(v) = \perp$. If, for all $y \in \{0,1\}^{t(|x|)}$, $v \neq i_{M,x}y$, then, by the definition of $D_M$, $D_M^{(0)}(v0^{t(|x|)+1}) = v0^{t(|x|)+1}$, $D_M(v0^{t(|x|)+1}) = v0^{t(|x|)}1$, and $D_M^{(0)}(v0^{t(|x|)}1) = v0^{t(|x|)}1$. Otherwise, let $y \in \{0,1\}^{t(|x|)}$ be such that $v = i_{M,x}y$. Then $D_M^{(0)}(v0^{t(|x|)+1}) = v0^{t(|x|)+1}$, $D_M^{(\nu(\hat{\theta}^{-1}(y))+1)}(v0^{t(|x|)+1}) = v0^{t(|x|)}1$, and $D_M^{(0)}(v0^{t(|x|)}1) = v0^{t(|x|)}1$.

For the induction case, suppose, for some $n$ that is less than the depth of $T_{M,x}$ and all subtrees $T$ of $T_{M,x}$ having depth at most $n$, that the induction hypothesis holds. Let $S$ be a subtree of $T_{M,x}$ of depth $n+1$, and let $v$ be the root of $S$. Let $\{a_1, \ldots, a_b\} = \mu_1^{-1}(v)$, where $a_b <_{\mathrm{lex}} \cdots <_{\mathrm{lex}} a_1$. It follows that each $a_1, \ldots, a_b$ is the root of a subtree of $S$ of depth at most $n$. By the definition of $D_M$, $D_M(v0^{t(|x|)+1}) = a_10^{t(|x|)+1}$, $D_M(a_10^{t(|x|)}1) = a_20^{t(|x|)+1}, \ldots, D_M(a_{b-1}0^{t(|x|)}1) = a_b0^{t(|x|)+1}$, and $D_M(a_b0^{t(|x|)}1) = v0^{t(|x|)}1$. By applying the induction hypothesis to the subtrees of $S$ rooted at $a_1, \ldots, a_b$, we conclude that $z$ is a node of $S$ if and only if there exist $i, j, k$ such that $D_M^{(i)}(v0^{t(|x|)+1}) = z0^{t(|x|)}1$, $D_M^{(j)}(z0^{t(|x|)+1}) = z0^{t(|x|)}1$, and $D_M^{(k)}(z0^{t(|x|)}1) = v0^{t(|x|)}1$.

Item 4 follows from the definition of $D_M$ (to see the case where $xwyza = x1^{2s(|x|)+t(|x|)-1}0^{t(|x|)+2}$, it helps to note that $\mu_1$ is undefined on $x1^{2s(|x|)+t(|x|)-1}0$ and thus $D_M(x1^{2s(|x|)+t(|x|)-1}0^{t(|x|)+2})$ is defined but $D_M(x1^{2s(|x|)+t(|x|)-1}0^{t(|x|)+1}1)$ is not).

For item 5, choose arbitrary $x \in \Sigma^*$, $w \in \{0,1\}^{2s(|x|)}$, $a \in \{0,1\}$, and $y, z \in \{0,1\}^{t(|x|)}$. If $xwyza = x1^{2s(|x|)+t(|x|)-1}0^{t(|x|)+2} \vee (wy \in \{0,1\}^{2s(|x|)+t(|x|)} - \{1^{2s(|x|)+t(|x|)-1}0\} \wedge z = 0^{t(|x|)})$ then, by item 3, there exists an $i \in \mathbb{N}$ such that $D_M^{(i)}(x1^{2s(|x|)+t(|x|)-1}0^{t(|x|)+2)}) = xwyza$. If $xw = i_{M,x} \wedge \nu(z) \leq \nu(\hat{\theta}^{-1}(y)) \wedge a = 0$ then, by the definition of $D_M$, $D_M^{(\nu(z))}(xwy0^{t(|x|)+1}) = xwyza$. Since, by item 3, there exists an $i \in \mathbb{N}$ such that $D_M^{(i)}(x1^{2s(|x|)+t(|x|)-1}0^{t(|x|)+2)}) = xwy0^{t(|x|)+1}$, it holds that $D_M^{(i+\nu(z))}(x1^{2s(|x|)+t(|x|)-1}0^{t(|x|)+2)}) = xwyza$.

For item 6, choose an arbitrary $x \in \Sigma^*$. Recall that, for all $xw \in \mathrm{R}_x - \{x1^m0^{2s(|x|)+t(|x|)-m}\}$, $\mathtt{dwn}(xw) \in \mathrm{R}_x$. Thus, since $x1^{2s(|x|)+t(|x|)} \in \mathrm{R}_x$ and $x1^{2s(|x|)+t(|x|)-1}0 \in \mathrm{R}_x$, it follows from the definitions of $\mathtt{dwn}$ and

$\leq_{\mathrm{R}_x}$ that, for some $i \in \mathbb{N}$, $\mathtt{dwn}^{(i)}(x1^{2s(|x|)+t(|x|)-1}0) = x1^{2s(|x|)+t(|x|)}$, and for all $j \in \mathbb{N}$ such that $0 \leq j \leq i$, it holds that $\mathtt{dwn}^{(j)}(x1^{2s(|x|)+t(|x|)-1}0) \in \mathrm{R}_x$. Thus, by the definition of $D_M$, $D_M^{(i)}(x1^{2s(|x|)+t(|x|)-1}0^{t(|x|)+2}) = x1^{2s(|x|)+t(|x|)}0^{t(|x|)+1}$, and for all $j \in \mathbb{N}$ such that $0 \leq j \leq i$, it holds that $D_M^{(j)}(x1^{2s(|x|)+t(|x|)-1}0^{t(|x|)+2}) = w0^{t(|x|)+1}$, where $w \in \mathrm{R}_x$.

For item 7, note that, by item 4, for all $w, y, z \in \Sigma^*$ such that $w \prec_{\mathrm{lex}} y \prec_{\mathrm{lex}} z$, either $D_M(y)$ is undefined or $D_M(z)$ is undefined. ❑

## Phase 5: Creating $A$

We are now ready to define $A$. $A$ is the same as the lexicographical ordering except that the strings between $x0^{2(s(|x|)+t(|x|))+1}$ and $x1^{2(s(|x|)+t(|x|))+1}$ are ordered as follows (let $z = x1^{2s(|x|)+t(|x|)-1}0^{t(|x|)+1}$).

- First come the strings $D_M^{(0)}(z0) = z0$, $D_M^{(1)}(z0) = D_M(z0)$, $D_M^{(2)}(z0), \ldots, z1$, in the order just stated.
- Next come the strings $\{xw \mid w \in \{0,1\}^{2(s(|x|)+t(|x|))+1} \wedge D_M(xw) \text{ is undefined} \wedge xw \neq z1\}$, in lexicographical order.

By Proposition 6.13.2, $A$ is a p-order. By Proposition 6.13.4, $A$ is total. By Propositions 6.13.1 and 6.13.7, $A$ has efficient adjacency checks.

**End of Construction**

We are now ready to prove Lemma 6.5.

*Proof of Lemma 6.5.* For each $f \in \mathrm{FPSPACE(poly)}$, we define $A$ as above. We define $b : \Sigma^* \to \Sigma^*$, $t : \Sigma^* \to \Sigma^*$, and $b' : \Sigma^* \to \Sigma^*$ on input $x \in \Sigma^*$ as, respectively, $b(x) =_{\mathrm{def}} x1^{2s(|x|)+t(|x|)}0^{t(|x|)+1}$, $t(x) =_{\mathrm{def}} x1^{2s(|x|)+t(|x|)}0^{t(|x|)}1$, and $b'(x) =_{\mathrm{def}} i_{M,x}y0^{t(|x|)+1}$, where $y = \theta(1)0^{t(|x|)-|\theta(1)|}$ (thus $\nu(\hat{\theta}^{-1}(y)) = 1$). Note that each of these functions is in FP.

For item 1, note that $s$ is polynomially bounded.

For item 2, we prove that, for all $x \in \Sigma^*$, $||\{z \mid b(x) <_A z <_A t(x)\}|| = 2^{2s(|x|)+1} + f(x) - 2$. Choose an arbitrary $x \in \Sigma^*$. By Proposition 6.13.4, both $D_M(x1^{2s(|x|)+t(|x|)}0^{t(|x|)+1})$ and $D_M(x1^{2s(|x|)+t(|x|)}0^{t(|x|)}1)$ are defined. Thus, by the definition of $A$, $\{z \mid b(x) <_A z <_A t(x)\} = \{z \mid (\exists i, k \in \mathbb{N} : i > 0 \wedge k > 0)[D_M^{(i)}(b(x)) = z \wedge D_M^{(k)}(z) = t(x)\}$. By Proposition 6.12.10, there are exactly $2^{2s(|x|)}$ strings in the subtree of $T_{M,x}$ rooted at $x1^{2s(|x|)+t(|x|)}$. Let $S = \{xwy0^{t(|x|)}a \mid w \in \{0,1\}^{2s(|x|)} \wedge a \in \{0,1\} \wedge y \in \{0,1\}^{t(|x|)} \wedge b(x) <_A xwy0^{t(|x|)}a <_A t(x)\}$. By Proposition 6.13.3, $||S|| = 2^{2s(|x|)+1} - 2$. By Propositions 6.12.6 and 6.13.3, there is a unique $y' \in \{0,1\}^{t(|x|)}$ such that $i_{M,x}y'0^{t(|x|)+1} \in \{z \mid b(x) <_A z <_A t(x)\}$. Moreover, by Proposition 6.12.8, $\nu(\hat{\theta}^{-1}(y')) = f(x)$. By the definition of $D_M$, $D_M^{(\nu(\hat{\theta}^{-1}(y'))+1)}(i_{M,x}y'0^{t(|x|)+1}) \in S$, and for each $i \in \mathbb{N}$ such that $0 < i \leq \nu(\hat{\theta}^{-1}(y'))$, it holds that $D_M^{(i)}(i_{M,x}y'0^{t(|x|)+1}) \notin S$. For each of the remaining $2^{2s(|x|)+1} - 3$ strings $w$ in $S$, $D_M(w) \in S \cup \{t(x)\}$. Thus $||\{z \mid b(x) <_A z <_A t(x)\}|| = 2^{2s(|x|)+1} + f(x) - 2$.

For item 3, we prove that $\|\{z \mid b'(x) <_A z <_A t(x)\}\| > 0$ if and only if $f(x) = 1$. Choose $x \in \Sigma^*$ and let $y = \theta(1)0^{t(|x|)-|\theta(1)|}$. Suppose that $f(x) = 1$. Then, by Proposition 6.12.8, $xi_{M,x}y$ is in the subtree of $T_{M,x}$ rooted at $x1^{2s(|x|)+t(|x|)}$. Thus, by Proposition 6.13.3, there exists a $k$ such that $D_M^{(k)}(b'(x)) = t(x)$. By the definitions of $D_M$, $b'$, and $t$, $D_M(b'(x)) \neq t(x)$, thus $k > 1$. By the definition of $A$, $\|\{z \mid b'(x) <_A z <_A t(x)\}\| > 0$. Now, suppose $f(x) \neq 1$. Since $f(x) \neq \nu(\hat{\theta}^{-1}(y))$, it follows from Proposition 6.12.8 that $i_{M,x}y$ is not in the subtree of $T_{M,x}$ rooted at $x1^{2s(|x|)+t(|x|)}$. Thus, by Proposition 6.13.3, for all $k \in \mathbb{N}$, $D_M^{(k)}(b'(x)) \neq t(x)$. Thus $b'(x) \not<_A t(x)$, and so $\|\{z \mid b'(x) <_A z <_A t(x)\}\| = 0$. ❑

## 7 The Complexity of Counting Divisors

Consider the function $\#\mathrm{DIV} : \mathbb{N} \to \mathbb{N}$, defined on input $m \in \mathbb{N}$ as

$$\#\mathrm{DIV}(m) =_{\mathrm{def}} \begin{cases} \|\{n \in \mathbb{N} \mid n \neq 1, n \neq m, \text{ and } n \text{ divides } m\}\| & \text{if } m \geq 1, \\ 0 & \text{otherwise.} \end{cases}$$

What can we say about its complexity? We claim that #DIV belongs to the interval size function class $\mathrm{IF_p}$.

**Theorem 7.1** *#DIV is in* $\mathrm{IF_p}$.

*Proof.* Let PRIMES be the set of all prime numbers. Observe that $\#\mathrm{DIV} \in \#\mathrm{P}$ and $\mathrm{PRIMES} = \{x \mid \#\mathrm{DIV}(x) = 0\}$. PRIMES $\in$ P [AKS02]. Thus Theorem 7.1 follows from Theorem 5.3. ❑

## 8 The Complexity of Counting Satisfying Assignments of Monotone Formulas

In this section, we show that the #MONSAT function fits into our collection of function classes. A *monotone boolean function* is any boolean function such that changing an input from 0 to 1 (while keeping all other inputs fixed) never changes the value of the function from 1 to 0. A *positive boolean formula* is a boolean formula that computes a monotone boolean formula. A *monotone boolean formula* is a formula having only $\wedge$ and $\vee$ connectors. Note that the class of functions computed by monotone boolean formulas is exactly the monotone boolean formulas. Monotone computing models have long been studied (see, e.g., Grigni and Sipser [GS92] and the references therein).

Define

$$\#\mathrm{MONSAT}(F) =_{\mathrm{def}} \begin{cases} \|\{(a_1, \ldots, a_n) \mid \\ \quad (\forall i : 1 \leq i \leq n)[a_i \in \{0,1\}] \wedge F(a_1, \ldots, a_n) = 1\}\| & \text{if } F \text{ is a monotone boolean formula} \\ 0 & \text{otherwise,} \end{cases}$$

i.e., $\#\mathrm{MONSAT}(F)$ counts the number of satisfying assignments of monotone boolean formulas. For the remainder of this section, we identify each assignment $(a_1, \ldots, a_n)$ to the $n$ variables of $F$ with the $n$-bit string $a_1 \ldots a_n \in \{0,1\}^n$. Theorem 8.5 states that #MONSAT belongs to the class $\mathrm{IF_t}$. To prove this theorem, we will use the following proposition.

**Proposition 8.1** *Let $\varphi$ be the function that is defined for every boolean formula $F(x_1, \ldots, x_n)$, $a \in \{0,1\}^n$, and $r \in \{0,1\}$ as $\varphi(F, a, r) =_{\mathrm{def}} \min\{\, b \mid b \in \{0,1\}^n \wedge a \leq_{\mathrm{lex}} b \wedge F(b) = r \,\}$ if $\{\, b \mid b \in \{0,1\}^n \wedge a \leq_{\mathrm{lex}} b \wedge F(b) = r \,\}$ is nonempty and $F$ is a monotone boolean formula, and $\varphi(F, a, r) =_{\mathrm{def}} \perp$ otherwise, where the* min *in the above definition is taken with respect to the lexicographical order. The function $\varphi$ is polynomial-time computable.*

*Proof.* To prove this proposition we use two natural properties of monotone boolean formulas. First, note that, for each monotone boolean formula $F$ of arity $n$ and for each $a = a_1 \ldots a_n \in \{0,1\}^n$ and $b = b_1 \ldots b_n \in \{0,1\}^n$, it holds that $F(a) \leq F(b)$ whenever $(\forall i \leq n)[a_i \leq b_i]$. Second, there is an assignment making $F$ true

```
[1]   b ← a
[2]   if (b = 1^n and F(b) ≠ r) or F(r^n) ≠ r
[3]      then
[4]         return ⊥
[5]      else
[6]         while b ≠ ε and F(b r^{n−|b|}) ≠ r do
[7]            b ← the string which succeeds b in lexicographical order
[8]            b ← longest prefix of b which ends with 1
[9]         endwhile
[10]        m ← |b| + 1
[11]        for j ← m to n do
[12]           if F(b0 r^{n−|b|−1}) = r
[13]              then
[14]                 b ← b0
[15]              else
[16]                 b ← b1
[17]           endif
[18]        endfor
[19]        return b
[20]  endif
```

Figure 6: An algorithm used in the proof of Proposition 8.1.

(respectively, false) if and only if $F(1^n) = 1$ (respectively, $F(0^n) = 0$). Consider the algorithm of Figure 6 running on an $n$-ary monotone boolean formula $F$, $a \in \{0,1\}^n$, and $r \in \{0,1\}$.

The algorithm works as follows. If none of the boundary conditions in lines 1–6 are met, then assume that the assignments to the variables of $F$ are just the labels of the leaves of a complete binary tree having $2^n$ leaves, i.e., the leftmost leaf is $0^n$, and the rightmost leaf $1^n$. The algorithm starts in the leaf numbered $a$, and searches the next node $u$ on the path from $a$ to the root such that the path comes into $u$ from the left, and the right subtree below $u$ contains an assignment $b$ with $F(b) = r$ (lines [6] to [9]). The least $b$ of the subtree having this property is determined via binary search (lines [10] to [18]). Thus, the algorithm is correct and runs in polynomial time with respect to the input length. ❑

We state as Proposition 8.2 some subcases of Proposition 8.1. (A "part 2 of Proposition 8.2" parallel to the first sentence of part 1 of Proposition 8.2 is not included since that trivially holds (test the all-0 assignment).) Though we could not find Proposition 8.2 in the literature, it is sufficiently fundamental that we believe it may well be known or a folk theorem.

**Proposition 8.2**

1. *The problem of finding the least satisfying assignment for monotone boolean formulas has a polynomial-time algorithm. Indeed, the problem of finding the least satisfying assignment lexicographically greater than or equal to a given assignment has, for monotone boolean formulas, a polynomial-time algorithm.*

2. *The problem of finding the least unsatisfying assignment lexicographically greater than or equal to a given assignment has, for monotone boolean formulas, a polynomial-time algorithm.*

This section has, so far, spoken of monotone boolean formulas. However, note that if we view the algorithm from Figure 6 as accessing a black-box boolean *function*, the algorithm in fact shows that the *query complexity* of the task is polynomial—indeed linear—if the black-box function is a monotone boolean function. Thus we have the following results.

**Proposition 8.3** *Let $\varphi$ be the function that is defined for every $n \geq 1$, every boolean formula $f(x_1, \ldots, x_n)$, every $a \in \{0,1\}^n$, and every $r \in \{0,1\}$ as*

$$\varphi^f(a,r) =_{\mathrm{def}} \begin{cases} \min\{b \mid b \in \{0,1\}^n \wedge a \leq_{\mathrm{lex}} b \wedge f(b) = r\} & \textit{if } \{b \mid b \in \{0,1\}^n \wedge a \leq_{\mathrm{lex}} b \wedge f(b) = r\} \neq \emptyset \\ \perp & \textit{otherwise,} \end{cases}$$

*where the* min *in the above definition is taken with respect to the lexicographical order. When restricted to monotone boolean functions, the function $\varphi$ is of linear (in the number of variables) query complexity (and polynomial, in the number of variables, time complexity). That is, there exist a Turing machine $M$ and a linear function $q$ and a polynomial $s$ such that for each $n \geq 1$, each monotone boolean $n$-variable function $f$, each $a \in \{0,1\}^n$, and each $r \in \{0,1\}$ it holds that*

1. *$M^f(a,r)$ makes at most $q(n)$ queries to $f$, and*
2. *$M^f(a,r)$ halts within $s(n)$ steps with $\varphi^f(a,r)$ on its output tape.*

Similarly to Proposition 8.2, we have the following (where the time and query complexities are relative to the number of variables (or, equivalently, relative to the size of the "input," i.e., $|a| + |r|$).

**Proposition 8.4**

1. *The problem of finding the least satisfying assignment when restricted to monotone boolean functions has a linear-query-complexity algorithm (that in addition is of polynomial-time complexity). Indeed, the problem of finding the least satisfying assignment lexicographically greater than or equal to a given assignment has, when restricted to monotone boolean functions, a linear-query-complexity algorithm (that in addition is of polynomial-time complexity).*
2. *The problem of finding the least unsatisfying assignment lexicographically greater than or equal to a given assignment has, when restricted to monotone boolean functions, a polynomial-time algorithm.*

Note that in neither Proposition 8.3 nor Proposition 8.4 do we make any claims about what the procedure will compute if the black-box function is not a monotone boolean formula.

We now relate #MONSAT to interval functions.

**Theorem 8.5** *#MONSAT $\in \mathrm{IF_t}$.*

*Proof.* We assume that $F$ is given as a string over the alphabet $\Sigma$. We construct a total p-order $A \in \mathrm{P}$ having efficient adjacency checks as follows. Generally, $A$ coincides with the lexicographical order on $\Sigma^*$ except that, for each monotone boolean formula $F$ of arity $n$, the interval between $1^{|F|}0F0000^n$ and $1^{|F|}0F1001^n$ is ordered in the following way.

- First comes $\{1^{|F|}0F000y \mid |y| = n\}$ in lexicographical order (we always use $n = n_F$ to denote the arity of $F$).
- Next comes the set $\{1^{|F|}0F001a \mid a$ is a satisfying assignment of $F\}$ in lexicographical order.
- Next comes $\{1^{|F|}0F010y \mid |y| = n\}$ in lexicographical order.

- Next comes the set $\{1^{|F|}0F011a \mid a \text{ is not a satisfying assignment of } F\}$ in lexicographical order.

- Finally comes the set $\{1^{|F|}0F100y \mid |y| = n\}$ in lexicographical order.

Clearly, $A$ is a total p-order that is decidable in polynomial time. In light of the function $\varphi$ from Proposition 8.1 it is not hard to see that $A$ has efficient adjacency checks. Also, for any monotone boolean formula $F(x_1, \ldots, x_n)$, let $b(F) =_{\text{def}} 1^{|F|}0F0001^n$ and $t(F) =_{\text{def}} 1^{|F|}0F0100^n$. Obviously, $b, t \in \text{FP}$, and we obtain $\#\text{MONSAT}(F) = \|\{z \mid b(F) <_A z <_A t(F)\}\|$. Thus, $\#\text{MONSAT} \in \text{IF}_{\text{t}}$. ❑

Valiant [Val79] showed that counting the number of satisfying assignments of 2CNF monotone formulas is Turing complete for #P. Since #2CNFMONSAT metrically reduces to #MONSAT, we immediately obtain from this theorem that #MONSAT is complete for $\text{IF}_{\text{t}}$ under Turing reductions, and we get an alternate proof for Corollary 5.8.

# 9 Cluster Computations

Finally, we discuss the complexity of computing the size of intervals for which the boundaries are not required to be polynomial-time computable. This leads to the notion of cluster computation, as introduced in [Kos99] for the case of the lexicographical order. We first review the formal definitions related to cluster computation, but here we present a more general version of the definitions than what previously appeared in [Kos99].

Let $M$ be any nondeterministic Turing machine that is "balanced" in the sense that, on every input, the graph of the nondeterministic choices $M$ makes is a complete, balanced, binary tree. Let $y$ and $z$ encode computation paths of $M$ on $x$. By the above assumption that $M$ is "balanced," $|y| = |z|$. Fix a total order $A$ on $\Sigma^*$. We say that $y \sim_{A,M,x} z$ if and only if (a) $y \prec_A z$ or $z \prec_A y$, and (b) $M$ on $x$ accepts on path $y$ if and only if $M$ on $x$ accepts on path $z$. Let $\equiv_{A,M,x}$ be the equivalence closure (i.e., the reflexive-symmetric-transitive closure) of $\sim_{A,M,x}$. Then the relation $\equiv_{A,M,x}$ is an equivalence relation and thus induces a partitioning of the computation tree of $M$ on $x$. An $A$-cluster is an equivalence class whose representatives are accepting paths.

For a nondeterministic Turing machine $M$, let $acc_M(x) \subseteq \Sigma^*$ denote the set of all accepting paths of $M$ on input $x$. Let $\#acc_M : \Sigma^* \to \mathbb{N}$ be the function defined as $\#acc_M(x) =_{\text{def}} \|acc_M(x)\|$. Let $out_M(x) \subseteq \Sigma^*$ denote the set of all distinct outputs of accepting paths of $M$ on input $x$. A nondeterministic Turing machine $M$ is a *lexicographical cluster machine* if and only if $M$ is balanced in the sense defined earlier and, for every $x$, there is a computation path $y$ of $M$ on $x$ such that

$$acc_M(x) = \{z \mid z \equiv_{\text{lex},M,x} y \text{ and } y \in acc_M(x)\}.$$

The intuition here is simple: Such machines on each input in the set have a single, nonempty, contiguous stretch of accepting paths.

**Definition 9.1** [Kos99]
c#P $=_{\text{def}} \{\#acc_M \mid M$ *is a polynomial-time lexicographical cluster machine*$\}$.

We mention some basic properties of the class c#P.

**Definition 9.2** *A nondeterministic Turing machine* computes a function $f$ almost-uniquely *if and only if, for each* $x$,

- $f(x) > 0$ *implies* $out_M(x) = \{f(x)\}$ *and* $\#acc_M(x) = 1$*, and*

- $f(x) = 0$ *implies* $out_M(x) = \emptyset$*.*

Recalling from Section 6.1 the definition of $\ominus$, we have the following.

**Proposition 9.3** [Kos99]

1. *A function* $f$ *lies in* c#P *if and only if there exists a nondeterministic polynomial-time Turing machine that computes* $f$ *almost-uniquely.*

2. $\mathrm{UPSV_t} \subseteq \mathrm{c\#P} = \mathrm{c\#P} \ominus \mathrm{FP} \subseteq \mathrm{\#P}$.

3. $\mathrm{UPSV_t} \cap \mathrm{Nonzero} = \mathrm{c\#P} \cap \mathrm{Nonzero}$.

4. $\mathrm{c\#P} = \mathrm{\#P}$ *if and only if* $\mathrm{UP} = \mathrm{PP}$.

**Proposition 9.4** 1. $\exists \cdot \mathrm{c\#P} = \exists \cdot (\mathrm{c\#P} - \mathrm{FP}) = \mathrm{UP}$.

2. *If* $\mathrm{IF_t} \subseteq \mathrm{c\#P}$ *then* $\mathrm{UP} = \mathrm{PP}$.

3. *If* $\mathrm{c\#P} \subseteq \mathrm{IF_t}$ *then* $\mathrm{P} = \mathrm{UP}$.

*Proof.* (1): It is easy to see that $\mathrm{UP} \subseteq \exists \cdot \mathrm{c\#P}$, since any balanced machine for a given UP language already implicitly shows that that language is in $\exists \cdot \mathrm{c\#P}$ due to the unique paths being each a size-one equivalence class. It follows from the definitions that $\exists \cdot (\mathrm{c\#P} \ominus \mathrm{FP}) \subseteq \exists \cdot (\mathrm{c\#P} - \mathrm{FP})$ and from Proposition 9.3.2 we have $\exists \cdot \mathrm{c\#P} = \exists \cdot (\mathrm{c\#P} \ominus \mathrm{FP})$. However, in light of Proposition 9.3.1, we can see that each set in $\exists \cdot (\mathrm{c\#P} - \mathrm{FP})$ is in fact in UP.

(2): By Theorem 5.7, $\mathrm{IF_t} \subseteq \mathrm{c\#P}$ implies $\mathrm{\#P} - \mathrm{FP} \subseteq \mathrm{c\#P} - \mathrm{FP}$. From this, Proposition 2.2.4, and the first part of the present result we have $\mathrm{PP} = \exists \cdot (\mathrm{\#P} - \mathrm{FP}) \subseteq \exists \cdot (\mathrm{c\#P} - \mathrm{FP}) = \mathrm{UP}$.

(3): Apply the operator $\exists$ to both sides of the inclusion, and apply Lemma 5.9 and the first part of the present result. ❑

Proposition 9.3, which in essence says that c#P functions are relatively simple, is *extremely* dependent on the fact that c#P is built based on lexicographical order. In particular, the results reflect the fact that it is easy, given two strings, $a$ and $b$, to compute $\|\{c \mid a \leq_{\mathrm{lex}} c \leq_{\mathrm{lex}} b\}\|$. Proposition 9.3.1 for example is driven in large part by the fact that one can, for inputs where the function is not zero, guess (and check the guess of) the rightmost and leftmost accepting paths, and then, since one knows that the complete set of accepting paths is simply the contiguous block between and including these, one can easily compute the number of accepting paths.

It is natural to wish to remove the focus here on lexicographic order, and to instead study machines whose set of accepting paths is always a contiguous block—with respect to some total order that has efficient adjacency checks like lexicographic order, but that perhaps does not satisfy the extremely restrictive "interval sizes are always trivial to compute" property of lexicographic order. We introduce the class CL#P, which captures exactly this more flexible, natural notion of cluster computing.

An order $A$ on $\Sigma^*$ is said to be *length-respecting* if and only if, for all $x, y$, $|x| < |y|$ implies $x <_A y$. Note that a length-respecting order is always a p-order.

**Definition 9.5** *A function $f$ belongs to the class* CL#P *if and only if there exist a nondeterministic polynomial-time Turing machine $M$, a polynomial $p$, and a length-respecting total order $A$ with efficient adjacency checks such that, for all $x$, the following conditions hold.*

1. *All computation paths of $M$ on $x$ have length exactly $p(|x|)$.*
2. *The set of all accepting paths of $M$ on $x$ is an $A$-cluster.*
3. *$f(x) = \#acc_M(x)$.*

As might be expected, the class $\mathrm{IF_t}$ is included in CL#P. Indeed, the following inclusions hold.

**Theorem 9.6** $\mathrm{c\#P} \cup \mathrm{IF_t} \subseteq \mathrm{CL\#P} \subseteq \mathrm{\#P}$.

*Proof.* The inclusions $\mathrm{c\#P} \subseteq \mathrm{CL\#P}$ and $\mathrm{CL\#P} \subseteq \mathrm{\#P}$ are trivial. It remains to prove the inclusion $\mathrm{IF_t} \subseteq \mathrm{CL\#P}$. Choose $f \in \mathrm{IF_t}$ via a total p-order $A \in \mathrm{P}$ having polynomial-time adjacency checks, functions $b, t \in \mathrm{FP}$, and a polynomial $p$ that witnesses that $A$ is a p-order. We may without loss of generality assume that $p$ is monotonic. For each $x \in \Sigma^*$, let $S_x = \{x0^{p(|x|)-|y|}1y0 \mid y \leq_A x\}$. Define $A'$ as follows. Generally, $A'$ corresponds to the lexicographical order on $\Sigma^*$, except that, for every $x \in \Sigma^*$, the interval between $x0^{p(|x|)+2}$ and $x1^{p(|x|)+2}$ is defined as follows.

- First come all strings in $S_x$, such that, for any strings $x0^{p(|x|)-|y_1|}1y_10, x0^{p(|x|)-|y_2|}1y_20 \in S_x$, let $x0^{p(|x|)-|y_1|}1y_10 \leq_{A'} x0^{p(|x|)-|y_2|}1y_20$ if and only if $y_1 \leq_A y_2$.
- Next come all the strings not in $S_x$, in lexicographical order.

We claim that $A'$ is a total, polynomial-time computable p-order having efficient adjacency checks. Clearly, $A'$ is total. Also, it is clear that, for any $s \in \Sigma^*$, it is possible to determine in polynomial time whether there is an $x \in \Sigma^*$ such that $s \in S_x$. It follows by this and by the definition of $A$ that $A'$ is polynomial-time computable. We claim that $A'$ has efficient adjacency checks. For any $x \in \Sigma^*$, the lexicographically smallest element in $S_x$ is $x0^{p(|x|)-|s_A|}1s_A0$, where $s_A \in \Sigma^*$ is the smallest element in the ordering imposed by $A$, and the lexicographically largest element is $x0^{p(|x|)-|x|}1x0$. If $x0^{p(|x|)-|y_1|}1y_10, x0^{p(|x|)-|y_2|}1y_20 \in S_x$ then $x0^{p(|x|)-|y_1|}1y_10 \prec_{A'} x0^{p(|x|)-|y_2|}1y_20$ if and only if $y_1 \prec_A y_2$ (this is true because, for every $y \in \Sigma^*$ such that $y \leq_A x$, it holds that $x0^{p(|x|)-|y|}1y0 \in S_x$; and thus, for such $y_1$ and $y_2$, it is impossible for some string longer than $p(|x_0|)$ to be "wedged between" them). The lexicographically smallest element not in $S_x$ is $x0^{p(|x|)+2}$ and the largest is $x1^{p(|x|)+2}$. For any $w_1, w_2 \in \Sigma^*$ and $b_1, b_2 \in \{0, 1\}$ such that both $w_1b_1$ and $w_2b_2$ are lexicographically between $x0^{p(|x|)+2}$ and $x1^{p(|x|)+2}$ but neither is in $S_x$, $w_1b_1 \prec_A w_2b_2$ iff $(w_1b_1 \prec_{\mathrm{lex}} w_2b_2)$ or $(w_1b_1 \not\prec_{\mathrm{lex}} w_2b_2$ and $b_1 = b_2 = 1$ and $w_1 \prec_{\mathrm{lex}} w_2$ and $w_20 \in S_x)$. All other cases are handled in the way obvious from the above, e.g., for any $w_1, w_2 \in \Sigma^*$ and $b_1, b_2 \in \{0, 1\}$ such that both of $w_1b_1$ and $w_2b_2$ are lexicographically between $x0^{p(|x|)+2}$ and $x1^{p(|x|)+2}$, and exactly one of them—say $w_1b_1$—is in $S_x$, the above makes it clear that $w_1b_1 \prec_{A'} w_2b_2$ exactly if $w_1b_1 = x0^{p(|x|)-|x|}1x0$ and $w_2b_2 = x0^{p(|x|)+2}$.

Define $M$ to be a Turing machine that, on input $x \in \Sigma^*$, guesses a string $w \in \Sigma^{p(|t(x)|)+2}$. If $t(x)w \notin S_{t(x)}$ then $M$ rejects. Otherwise, $M$ accepts iff $t(x)0^{p(|t(x)|)-|b(x)|}1b(x)0 <_{A'} t(x)w <_{A'} t(x)0^{p(|t(x)|)-|t(x)|}1t(x)0$. Clearly, $M$ runs in polynomial time and has computation paths of length exactly $p(t(|x|)) + 2$. Also, the number of accepting paths of $M$ on $x$ equals $f(x)$. By construction, the set of accepting computation paths of $M$ on $x$ is an $A'$-cluster. Thus, $f \in \mathrm{CL\#P}$. ❑

From Proposition 9.4 and Theorem 9.6, it is clear that CL#P is different from both c#P and $\mathrm{IF_t}$ unless some surprising complexity class collapses occur. In particular, the following holds.

**Corollary 9.7** *1. If* c#P = CL#P*, then* UP = PP.

*2. If* $\mathrm{IF_t}$ = CL#P*, then* P = UP.

Nonetheless, when considering only polynomially bounded functions, c#P and CL#P do coincide.

**Theorem 9.8** c#P ∩ PolyBounded = CL#P ∩ PolyBounded.

*Proof.* The inclusion "⊆" is immediate. For the inclusion "⊇," choose $f \in$ CL#P via a nondeterministic polynomial-time Turing machine $M$, a polynomial $p$, and a length-respecting total order $A$ having efficient adjacency checks, all three of which have the properties and behaviors described in Definition 9.5. Recall that all accepting paths of $M$ on any input $x$ will be of length $p(|x|)$. Let $q$ be a polynomial such that, for all $x \in \Sigma^*$, $f(x) \leq q(|x|)$. We now will define a nondeterministic polynomial-time Turing machine $N$ that almost-uniquely computes $f$ in the sense of Definition 9.2. Define $N$ to be a Turing machine that, on input $x \in \Sigma^*$, does the following.

1. If $\epsilon$ is an accepting path of $M(x)$ then accept and output 1.

2. $N$ nondeterministically guesses strings $y, z \in \Sigma^{p(|x|)}$, $y' \in \Sigma^{p(|x|)-1} \cup \Sigma^{p(|x|)}$, and $z' \in \Sigma^{p(|x|)} \cup \Sigma^{p(|x|)+1}$.

3. $N$ checks whether all of the following hold.

   (a) $y' \prec_A y$ and $z \prec_A z'$.

   (b) $y' \notin acc_M(x)$.

   (c) $z' \notin acc_M(x)$.

   (d) $y \in acc_M(x)$ and $z \in acc_M(x)$.

4. If (3) does not hold, then $N$ rejects, otherwise if $y = z$, $N$ accepts and outputs 1.

5. If (3) does hold and $y \neq z$, then $N$ proceeds as follows.

   (a) $N$ nondeterministically guesses an integer $r$ with $0 \leq r \leq q(|x|) - 2$.

   (b) $N$ nondeterministically guesses $r$ strings $v_1, \ldots, v_r \in \Sigma^{p(|x|)}$.

   (c) $N$ checks whether $y \prec_A v_1 \prec_A v_2 \prec_A \cdots \prec_A v_r \prec_A z$.

   (d) If (5c) does not hold, then $N$ rejects. Otherwise, $N$ accepts and outputs $r + 2$.

$N$ is a nondeterministic polynomial-time Turing machine that, on each input, has one accepting path if $f(x) > 0$ and no accepting paths if $f(x) = 0$. If $f(x) > 0$, then $N$ on $x$ outputs $f(x)$ on its accepting path. Thus, $N$ almost-uniquely computes $f$, and so by Proposition 9.3.1 $f \in$ c#P. ❑

For a class $\mathcal{F}$ of functions, let $\exists! \cdot \mathcal{F}$ be the class of all sets $L$ for which there exists a function $f \in \mathcal{F}$ such that, for all $x$, $x \in L \Leftrightarrow f(x) = 1$.

**Theorem 9.9** *1.* $\exists! \cdot \mathrm{IF_p}$ = coNP.

*2.* $\exists! \cdot \mathrm{c\#P} = \exists! \cdot \mathrm{CL\#P} = \mathrm{UP}$.

*Proof.* For (1), $\mathrm{coNP} \subseteq \exists! \cdot \mathrm{IF_p}$ follows from Corollary 5.4 and the observation that any language in coNP is also (via considering the NP machine for the language's complement but with one extra accepting path added on each input) in $\exists! \cdot (\#\mathrm{P} \cap \mathrm{Nonzero})$. To see $\exists! \cdot \mathrm{IF_p} \subseteq \mathrm{coNP}$, choose $L \in \exists! \cdot \mathrm{IF_p}$, via $f \in \mathrm{IF_p}$. Let boundary functions $b, t \in \mathrm{FP}$ and partial, polynomial-time computable p-order $A$ having efficient adjacency checks witness that $f \in \mathrm{IF_p}$. Let $M$ be a nondeterministic polynomial-time Turing machine that, on input $x$, (i) guesses $y, z \in \Sigma^*$ such that $y \neq z$ and (ii) accepts if $b(x) \prec_A t(x) \vee (b(x) <_A y <_A t(x) \wedge b(x) <_A z <_A t(x))$. It is easy to see that $M$ accepts $\overline{L}$, thus $L \in \mathrm{NP}$.

For (2), $\mathrm{UP} \subseteq \exists! \cdot \mathrm{c\#P}$ is obvious. To see that $\exists! \cdot \mathrm{CL\#P} \subseteq \mathrm{UP}$, choose $L \in \exists! \cdot \mathrm{CL\#P}$. Thus there exists a function $f \in \mathrm{CL\#P}$ such that, for all $x$, $x \in L \Leftrightarrow f(x) = 1$. Let $M$ be a machine that computes $f$ via total order $A$ having efficient adjacency checks and polynomial $p$ (where $M$, $A$, and $p$ are in the sense of Definition 9.5). Recall that all accepting paths of $M(x)$ are of length $p(x)$. Let $N$ be a nondeterministic polynomial-time Turing machine that, on input $x$, guesses strings $y \in \Sigma^{p(|x|)}$ and $x, z \in \Sigma^{p(|x|)-1} \cup \Sigma^{p(|x|)} \cup \Sigma^{p(|x|)+1}$, and accepts if and only if all the following hold.

1. $y \prec_A z \wedge y \in acc_M(x) \wedge (w \prec_A y \vee w = y = \epsilon)$.
2. $w \notin acc_M(x) \vee w = y = \epsilon$.
3. $z \notin acc_M(x)$.

Clearly, $N$ has on any input at most one accepting path and $N$ accepts $L$. ❑

The next result shows that CL#P is probably not powerful enough to capture #P.

**Theorem 9.10** *If* $\mathrm{CL\#P} = \#\mathrm{P}$ *then* $\mathrm{UP} = \mathrm{PH}$.

*Proof.* Using Theorem 5.2 and both parts of Theorem 9.9, we have $\mathrm{coNP} \subseteq \exists! \cdot \#\mathrm{P} = \exists! \cdot \mathrm{CL\#P} = \mathrm{UP}$. ❑

On the other hand, proving CL#P to be different from #P is at least as hard as proving that $\mathrm{P} \neq \mathrm{NP}$ and $\mathrm{UP} \neq \mathrm{PP}$.

**Proposition 9.11** *If* $\mathrm{P} = \mathrm{NP}$ *or* $\mathrm{UP} = \mathrm{PP}$ *then* $\mathrm{CL\#P} = \#\mathrm{P}$.

*Proof.* Suppose $\mathrm{UP} = \mathrm{PP}$. Then by Proposition 9.3.4 $\mathrm{c\#P} = \#\mathrm{P}$, and so (see Theorem 9.6) $\mathrm{CL\#P} = \#\mathrm{P}$. Suppose that $\mathrm{P} = \mathrm{NP}$. Then by Theorem 5.10 it holds that $\mathrm{IF_t} = \#\mathrm{P}$, and so (see Theorem 9.6) $\mathrm{CL\#P} = \#\mathrm{P}$. ❑

Unfortunately, the necessary and sufficient conditions we have obtained for the equality of #P and CL#P differ, i.e., they do not yield a complete characterization. However, if we consider polynomially bounded functions, then such a complete characterization can be established in terms of the classes UP [Val76] and Few [CH90] (see Section 2 for a review of their definitions). Note that $\mathrm{UP} = \mathrm{Few} \Leftrightarrow \mathrm{UP} = \mathrm{coUP} = \mathrm{FewP} = \mathrm{Few}$ and so in light of Theorem 9.12 we easily have that $\mathrm{CL\#P} \cap \mathrm{PolyBounded} = \#\mathrm{P} \cap \mathrm{PolyBounded}$ implies $\mathrm{UP} = \mathrm{coUP} = \mathrm{FewP}$.

**Theorem 9.12** $\mathrm{CL\#P} \cap \mathrm{PolyBounded} = \#\mathrm{P} \cap \mathrm{PolyBounded}$ *if and only if* $\mathrm{UP} = \mathrm{Few}$.

*Proof.* [⇒]: Suppose that $L \in \text{Few}$ via a function $f \in \#\text{P}$, a set $B \in \text{P}$, and a polynomial $p$ such that, for all $x$, $f(x) \leq p(|x|)$, and $x \in L \Leftrightarrow (x, 1^{f(x)}) \in B$. Let $g(x) =_{\text{def}} 1 + f(x)$. Then $g \in \#\text{P}$, and $g$ is polynomially bounded. From our hypothesis and Theorem 9.8, we obtain $g \in \text{c}\#\text{P}$. Since $g(x) > 0$, by Theorem 9.3.3 we have that $g \in \text{UPSV}_{\text{t}}$ via some nondeterministic polynomial-time (function-computing) Turing machine $M$ whose behavior is $\text{UPSV}_{\text{t}}$-like. Define $N$ to be a Turing machine that, on input $x$, nondeterministically guesses a computation path $y$ of $M$ on input $x$, simulates $M$ on input $x$ along computation path $y$, and accepts (on its current path) if and only if $y$ is an accepting path with output $z$ satisfying $(x, 1^{z-1}) \in B$. Clearly, $N$ is a nondeterministic polynomial-time Turing machine with at most one accepting path on each input. Furthermore, it holds that $N$ on $x$ has an accepting computation path if and only if $(x, 1^{f(x)}) \in B$. This gives $L \in \text{UP}$.

[⇐]: Let $f$ be any polynomially bounded #P function. Define $A =_{\text{def}} \{(x, 1^y) \mid y \leq f(x)\}$. Note that $A \in \text{Few}$. So by our hypothesis $A \in \text{UP}$. Indeed, since Few is closed under complementation and Few = UP by hypothesis, $A \in \text{UP} \cap \text{coUP}$. Via binary search using $A$ as an oracle, we can compute $f$ in polynomial time. That is, $f$ is in $\text{FP}^{\text{UP} \cap \text{coUP}} = \text{UPSV}_{\text{t}} \subseteq \text{c}\#\text{P}$. Thus, $\text{CL}\#\text{P} \cap \text{PolyBounded} = \#\text{P} \cap \text{PolyBounded}$. ❑

From Corollary 9.7, we know that CL#P and c#P probably are different classes. However, under the $\exists$ operator the difference disappears, since both are mapped to UP. (Recall that Proposition 9.4.1 established $\exists \cdot \text{c}\#\text{P} = \text{UP}$.)

**Theorem 9.13** $\exists \cdot \text{CL}\#\text{P} = \text{UP}$.

*Proof.* The inclusion $\text{UP} \subseteq \exists \cdot \text{CL}\#\text{P}$ is immediate from Proposition 9.4.1 and the fact that $\text{c}\#\text{P} \subseteq \text{CL}\#\text{P}$. To show the inclusion $\exists \cdot \text{CL}\#\text{P} \subseteq \text{UP}$, choose an arbitrary $L \in \exists \cdot \text{CL}\#\text{P}$. Let $L \in \exists \cdot \text{CL}\#\text{P}$ via some function $f \in \text{CL}\#\text{P}$ with $x \in L \Leftrightarrow f(x) > 0$. Let $f \in \text{CL}\#\text{P}$ be witnessed (in the sense of the $M$, $p$, and $A$ of Definition 9.5) by some Turing machine $M$, polynomial $p$, and total order $A$ with efficient adjacency checks. Define $N$ to be a Turing machine that, on input $x \in \Sigma^*$, does the following.

1. $N$ nondeterministically guesses $z \in \Sigma^{p(|x|)}$ and $z' \in \Sigma^{p(|x|)} \cup \Sigma^{p(|x|)+1}$.
2. $N$ checks whether each of the following conditions holds.
   (a) $z \prec_A z'$.
   (b) $z \in acc_M(x)$.
   (c) $z' \notin acc_M(x)$.
3. $N$ accepts if and only if 2 holds.

Clearly, $N$ runs in polynomial time and always has at most one accepting path. Also, it holds that $\#acc_N(x) = 1 \Leftrightarrow x \in L$. Thus, $L \in \text{UP}$. ❑

It is known that c#P is not closed under increment unless UP = coUP [Kos99]. We note that CL#P displays the same behavior.

**Theorem 9.14** *If* CL#P *is closed under increment, then* UP = coUP.

*Proof.* Observe that $\text{co}(\exists \cdot \mathcal{F}) \subseteq \exists! \cdot (\mathcal{F} + 1)$ is true for every class $\mathcal{F}$ of total functions, where $\mathcal{F} + 1$ denotes $\{g \mid (\exists f \in \mathcal{F})(\forall x)[g(x) = f(x) + 1]\}$. Thus by our hypothesis and Theorem 9.13 we have $\text{coUP} = \text{co}(\exists \cdot \text{CL}\#\text{P}) \subseteq \exists! \cdot (\text{CL}\#\text{P} + 1) \subseteq \exists! \cdot \text{CL}\#\text{P} = \text{UP}$. ❑

As a corollary, we obtain that CL#P is incomparable to $\mathrm{IF_p}$ unless some unexpected complexity class collapse occurs.

**Corollary 9.15** *1. If* $\mathrm{CL\#P} \subseteq \mathrm{IF_p}$*, then* $\mathrm{P} = \mathrm{UP}$*.*

*2. If* $\mathrm{IF_p} \subseteq \mathrm{CL\#P}$*, then* $\mathrm{UP} = \mathrm{PH}$*.*

*Proof.* Regarding (1), from our hypothesis and Theorem 9.13 we have $\mathrm{UP} = \exists \cdot \mathrm{CL\#P} \subseteq \exists \cdot \mathrm{IF_p} = \mathrm{P}$. To verify (2), observe that from our hypothesis, Theorem 9.9.1, and Theorem 9.13 we obtain $\mathrm{coNP} \subseteq \exists! \cdot \mathrm{IF_p} \subseteq \exists! \cdot \mathrm{CL\#P} = \mathrm{UP}$. ❑

# 10 Conclusion and Open Problems

We introduced interval size functions over p-orders and used them to provide an alternate definition of #P as the set of all interval size functions over polynomial-time decidable p-orders. We also introduced the classes $\mathrm{IF_p}$ and $\mathrm{IF_t}$, the interval size functions over partial and total polynomial-time computable p-orders with efficient adjacency checks. We proved that $\mathrm{IF_p}$ is the class of all functions in #P whose support is in P. We also proved that $\mathrm{IF_t} - \mathrm{FP} = \#\mathrm{P} - \mathrm{FP}$ and $\mathrm{IF_p} - \mathcal{O}(1) = \#\mathrm{P} - \mathcal{O}(1)$, but that $\mathrm{IF_p} = \#\mathrm{P}$ if and only if $\mathrm{P} = \mathrm{NP}$, and that $\mathrm{IF_t} = \mathrm{IF_p}$ only if $\mathrm{UP} = \mathrm{PH}$.

We also introduced the classes $\mathrm{IF_p^*}$ and $\mathrm{IF_t^*}$, the interval size functions over partial and total p-orders with efficient adjacency checks. We proved that $\exists \cdot \mathrm{IF_t^*} = \exists \cdot \mathrm{IF_t^*} = \mathrm{PSPACE}$.

Finally, we introduced CL#P, the set of all functions that count the number of accepting paths of polynomial-time cluster machines whose underlying orders are total and have efficient adjacency checks, and we studied the relationship between CL#P and the previously-studied cluster computing class c#P.

Reviewing all the results on the interval size function classes $\mathrm{IF_p}$, $\mathrm{IF_p^*}$, $\mathrm{IF_t}$, and $\mathrm{IF_t^*}$, it seems that we have a good understanding of the computational power of the classes $\mathrm{IF_p}$, $\mathrm{IF_p^*}$, and $\mathrm{IF_t^*}$. Regarding the class $\mathrm{IF_t}$, we commend as an open issue obtaining an understanding of the class $\mathrm{IF_t} - \mathcal{O}(1)$, which can be loosely considered to be a kind of "total order" #P.

# Acknowledgments

We are grateful to J. Rothe, H. Spakowski, and M. Thakur for proofreading an earlier draft of this paper, and to E. Hemaspaandra and K.-J. Lange for helpful discussions.